\documentclass{aa}

\newcommand{\mycomment}[1]{}
\usepackage{graphicx}
\usepackage{xcolor}
\usepackage{longtable}

\usepackage{txfonts}

\usepackage{capt-of}
\usepackage{subfig}
\usepackage{amsmath}
\usepackage{booktabs}
\usepackage{multirow}

\usepackage[colorlinks,allcolors=blue]{hyperref}
\usepackage{orcidlink}  

\newcommand{\emcee}{\texttt{EMCEE}}
\newcommand{\emp}{\texttt{EMPEROR}}
\newcommand{\exo}{\texttt{EXOSTRIKER}}
\newcommand{\ariadne}{\texttt{ARIADNE}}
\newcommand{\species}{\texttt{SPECIES}}

\begin{document} 

   \title{Destruction of "peas in a pod?" A candidate multi-planet system around the nearby, bright star, HD208487}

   \author{Rafael I. Rubenstein\orcidlink{0000-0002-2460-0305}
          \inst{1}\fnmsep \inst{2}
          \and
          James S. Jenkins\orcidlink{0000-0003-2733-8725}\inst{1,3}
          \and
          Pablo A. Peña R.\orcidlink{0000-0002-8770-4398}\inst{1,3}
          \and
          Carolina Charalambous \inst{4,5}
          \and
          Mikko Tuomi \inst{6,7}
          \and
          Douglas R. Alves \inst{8}
          \and
          José Vines \inst{9}
          \and 
          Mat\'ias R. D\'iaz \inst{10}
          \and
          Suman Saha \inst{1,3}
          \and 
          R. Paul Butler \inst{11}
          \and 
          Jeffrey D. Crane \inst{12}
          \and 
          Steve Shectman \inst{12}
          \and 
          Johanna K. Teske \inst{11}
          \and 
          David Osip \inst{13}
          \and
          Zahra~Essack \inst{14}
          \and
          Benjamin~T.~Montet \inst{15,16}
          \and
          Adina~D.~Feinstein \inst{17,18,19}
          \and
          Cristobal Petrovich \inst{4,5,20}          
          }

    \titlerunning{Destruction of "peas in a pod?" A candidate multi-planet system around HD208487}

   \institute{Instituto de Estudios Astrofísicos, Universidad Diego Portales - Av. Ejército Libertador 441, Santiago, 8370191                    
   \and
   United States Fulbright Fellow; Chile Fulbright Commission; \email{Rafael.Rubenstein@fulbrightmail.org}
        \and
            Centro de Astrof\'isica y Tecnolog\'ias Afines (CATA), Casilla 36-D, Santiago, Chile
        \and
            Instituto de Astrof\'isica, Pontificia Universidad Cat\'olica de Chile, Av. Vicu\~na Mackenna 4860, 782-0436 Macul, Santiago, Chile
        \and
            Millennium Institute for Astrophysics (MAS), Chile
         \and
            University of Helsinki, Department of Physics, PO Box 64, 00014 Helsinki, Finland
        \and
            University of Hertfordshire, Center for Astrophysics, College Lane Campus, Hatfield, Hertfordshire, UK, AL10 9AB
        \and
            Departamento de Astronomía, Universidad de Chile, Casilla 36-D, Santiago, Chile
        \and 
            Instituto de Astronom\'ia, Universidad Cat\'olica del Norte, Angamos 0610, 1270709, Antofagasta, Chile
        \and
            Las Campanas Observatory, Carnegie Institution of Washington, Ra\'ul Bitr\'an 1200, La Serena, Chile.
        \and
            Earth and Planets Laboratory, Carnegie Institution for Science, 5241 Broad Branch Rd NW, Washington, DC 20015
        \and
            The Observatories of the Carnegie Institution for Science, 813 Santa Barbara Street, Pasadena, CA, 91101
        \and
            Las Campanas Observatory, Carnegie Institution for Science, Colina el Pino, Casilla 601 La Serena, Chile
        \and
            Department of Physics and Astronomy, The University of New Mexico, 210 Yale Blvd NE, Albuquerque, NM 87106, USA
        \and
            School of Physics, University of New South Wales, Sydney, NSW 2052, Australia
        \and
            UNSW Data Science Hub, University of New South Wales, Sydney, NSW 2052, Australia
        \and
            NHFP Sagan Fellow
        \and
            Laboratory for Atmospheric and Space Physics, University of Colorado Boulder, UCB 600, Boulder, CO 80309
        \and
            Department of Physics and Astronomy, Michigan State University, East Lansing, MI 48824, USA
        \and
            Department of Astronomy, Indiana University, Bloomington, IN 47405, USA
             }

   \date{Received February 9, 2024; accepted August 14, 2025}
 
  \abstract
   {}  
   {We reinvestigate the HD208487 star system to test the reality of the proposed HD208487c world. We also search for additional companions using applied Bayesian statistics and 15+ years of new radial velocity (RV) data from the HARPS and the PFS instruments that were taken post-discovery of HD208487b, as part of our continued study of bright Sun-like stars within 50~pc of the Sun.}
   {RV data were analyzed with Generalized Lomb-Scargle Periodograms, followed by Bayesian analysis techniques using the \emp\ code. 
   We also scrutinized various stellar activity indices to search for any corresponding peaks in the power spectra, correlations with the RV measurements, or significant signals from a Bayesian analysis methodology. Finally, photometric data were also checked to test for any transits or possible activity manifestations that could lead to possible false RV signals or excess noise.}
   {Our analysis points towards a candidate second planet in the system, positioned near the period of a previously proposed and subsequently challenged signal. This signal, HD208487c, would relate to a cool Saturn world with an orbital period of $ 923.06 {\genfrac{}{}{0pt}{}{+2.02}{-2.76}}d$ and a minimum mass of $M_jsini = 0.32 \pm 0.01 M_j$.  Our analysis also gives rise to a newly discovered candidate planet, HD208487d, which if confirmed would be the result of a cool super-Neptune/sub-Saturn with a period of $1380.13 {\genfrac{}{}{0pt}{}{+19.20}{-8.25}}d$ and a minimum mass of $M_jsini = 0.15 \pm 0.01 M_j$.  Neither stellar activity indices nor photometric data show signals statistically matching these periods. We find that stellar activity is indeed affecting the RVs, yet our joint RV+activity indicator modeling argues they are Doppler in nature. We show that the RV models are stable over long timescales, and these signals are independent of wavelength-dependent noise. The relative contributions of the data to the model were also examined.}
   {We uncovered a candidate three-planet system that would consist of an inner gas giant, a central Saturn, and an outer super-Neptune/sub-Saturn. Extensive analysis of both photometric and spectroscopic data as activity proxies strongly supports the planetary system hypothesis, yet more long-term RV data would help add more statistical weight to the reality of candidate planets c and d. Assuming our model best represents reality, a dynamical analysis suggests that gravitational scattering of an initially ordered, equally spaced system in a long resonant chain of six Neptunes can explain the current architecture of HD208487- a moderately-eccentric, inner massive planet with an outer nearly resonant ($P_d/P_c = 1.495$) gas giant and super-Neptune. More RVs may also shed light on the reality of a fourth Doppler signal uncovered in the data that sits close to the 2:1 period ratio with signal of HD208487c.}

   \maketitle

\section{Introduction}
    
In 1995, Mayor and Queloz discovered 51 Peg b, the first confirmed exoplanet around a main sequence star, using the Radial Velocity (RV) method \citep{51Peg}. This planetary detection represented a paradigm shift and launched the exoplanet field on the path towards its current sophisticated state. Specifically, their discovery proved giant planets could be detected with the technology at the time, resulting in a huge push to find more new worlds, such as 70 Vir b \citep{70Virginis}, and 47 UMa b \citep{47UrsaeMajoris}. Subsequently, more advanced telescopes with much higher resolution and precision were developed, allowing the detection of less massive planets, including, for example, the super-Earths GJ667C b, c, and d \citep{GJ667, GJ667_Revisited}. Additionally, astronomers searched for planets in the Habitable Zone (HZ), where liquid water could exist on a planet's surface. \cite{HD28185_Discovery} discovered HD28185, which \cite{Habitability208487} confirmed to be the first planet entirely within the HZ. Eventually, \cite{Kepler186f} discovered Kepler 186f, the first (of many) `Earth-like' planets orbiting its star in the habitable zone. As more and more discoveries accumulated, multi-planetary solar systems were discovered, including Upsilon Andromodae b, c, and d \citep{ThreeUpsilonAndromedae}. Further, there have now been many detections of multi-planetary systems in which two or more planets are locked in a mean-motion orbital resonance (MMR) \citep{OrbitalResonances, Migration2/1, Migration_21_32_Resonance}, such as the discovery of TOI-216b and c, two warm, large exoplanets in a 2:1 MMR \citep{TOI216_WarmResonance}. \\
   \\
   As the body of planetary systems with confirmed or candidate planets increases, it remains extremely important to revisit these systems to ensure the most accurate planetary parameters are recorded. It is not uncommon for a planet detection to later be contested due to future analyses with more data or differing methods producing inconsistent results \citep{jones14, Testingeccentric_multiples, Trulyeccentric2, Trulyeccentric1,jenkins2013,jenkins2014,Santos_stellaractivity}. For example, it is well-known that stellar activity can create periodic RV signals \citep{StellarActivity_RVJitter, LongtermStellarActivity_RVvariations}. Magnetic activity, photometric variations, stellar rotation cycles, or long-term magnetic cycles can imitate planetary RV signals \citep{MagCycles_ActivityVariation, MagneticCycles_Lovis}. Therefore, examining available stellar indices, photometric data, and any correlations with the RV data remains an important step in verifying true planetary RV signals \citep[e.g. see][]{HD26965Diaz}. On the other hand, re-examinations of previously studied star systems frequently lead to new planetary detections \citep{ HD40307_original,Gliese581, HD40307_reexamine, HD19193_Revisited_HZ}.\\
   \\
   The scientific community places a special emphasis on the search for exoplanets in nearby solar systems, a case in point being the discovery of Proxima b, a terrestrial world orbiting within the HZ of our nearest neighbour star, Proxima Centauri \citep{ProximaCentaurib}, since these offer immense hope to advance our understanding of planet formation and evolution. HD208487 serves as a particularly interesting system to re-examine, since it is one of the nearest ($\sim45$pc) and brightest ($V = 7.47$) stars in the southern sky, and it already is known to be orbited by one planet. Additionally, a second, longer period signal was detected \citep{Gregory_longerP, Gregoryupdated}, though later disregarded as likely activity-induced \citep{HD208487_neptune_14/998, HD208487_28d}.  We now have 184 new RV observations spanning more than 15 years since \cite{TinneyDiscoveryHD208487} published this original discovery of HD208487b. \\
\\
    Of the nearly 900 multiplanetary systems\footnote{The Extrasolar Planets Encyclopaedia (\href{http://exoplanet.eu/catalog}{http://exoplanet.eu})}, only a handful of these exhibit similarities to our own Solar System. 
   \cite{2018AJ....155...48W} analysed the architectures of the Kepler sample of multi-planet systems, unveiling a phenomenon denoted as `peas in a pod.' Within a multi-planet system, each planet's size is strongly correlated with that of its neighboring planets. In systems boasting three or more transiting planets, an additional pattern emerges: a systematic regularity in the spacing of planets. Frequently, these systems are densely populated with planets near their stability thresholds. Extensive research on the dynamic evolution and stability indicates that gravitational scattering plays a crucial role in shaping the final architecture of such multi-planetary systems \citep[e.g.,][]{2008ApJ...686..580C,2012ApJ...758...39J,2017MNRAS.470.1750I,2023AJ....166..267W,2024MNRAS.527...79G}. This process triggers perturbations in planetary orbits, giving rise to close encounters between planets.
   Ultimately, this can lead to either collisions (whether between two planets or between a planet and its star) or the ejection of planets from the system. The former is more prevalent at smaller semi-major axes \citep[see, e.g.,][]{2014ApJ...786..101P}. This observed correlation between planet size and spacing underscores the pivotal role of dynamics in shaping the final sizes and orbital separations of planets within these systems. \\
\\
   In this paper, we will discuss how our new analysis has lead to the statistical confirmation and discovery of two new planets in this system, HD208487c and HD208487d. The rest of the paper will follow the ensuing structure: Sect. 2 will discuss the RV observations in detail, including a breakdown of the measurements by telescope. In Sect. 3 we present updated stellar parameters for HD208487 using the \species\ and \ariadne\ codes. We begin the analysis of the RV dataset by observing the GLS periodograms in Sect. 4, while Sect. 5 details the detection of the three planets with the more rigorous Bayesian analysis. Next, we highlight the search for correlations between stellar activity indices and the RVs in Sect. 6. Sect. 7 describes the analysis of ASAS photometric data and the fitting of stellar cycles. Finally, we discuss the results and their implications in Sect. 8, followed by the conclusions in Sect. 9. 
 
\section{RV observations}

\begin{figure}[t]
   \centering
   \includegraphics[width=\columnwidth]{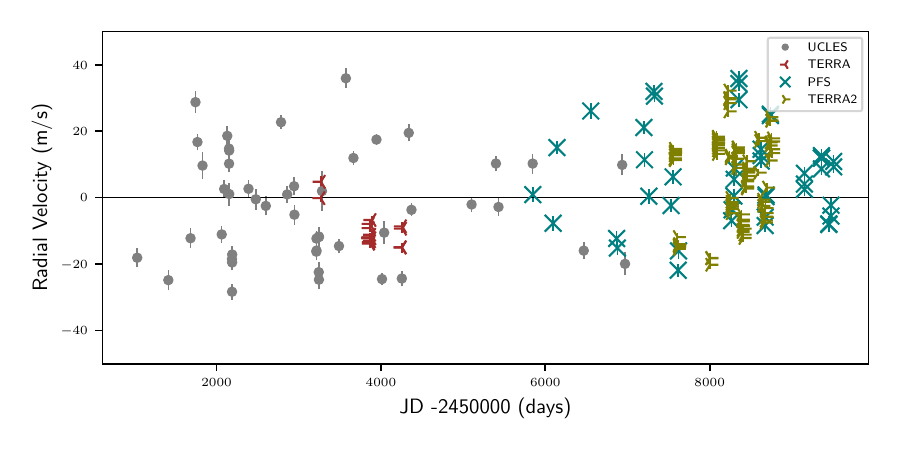}
      \caption{All 219 mean-subtracted radial velocity measurements of HD208487, including 48 UCLES (grey), 20 HARPS-TERRA (red), 106 HARPS-TERRA2 (olive), and 45 PFS (teal) observations.}
         \label{FullRVs}
   \end{figure}

    We used a total of 219 high-precision Doppler spectroscopic RV measurements of HD208487 spanning from JD 2451034.18, August 8, 1998, to JD 2459508.64, October 21, 2021 for a total baseline of 8474 days (23.2 years), as seen in Figure \ref{FullRVs}. The data comes from the University College London Échelle Spectrograph (UCLES) instrument installed on the 3.9m Anglo-Australian Telescope at the Siding Spring Observatory (SSO), the High Accuracy Radial Velocity Planet Searcher (HARPS) installed on the 3.6m ESO Telescope (before and after the fibre upgrade in 2015) at La Silla Observatory, and the Carnegie Planet Finder Spectograph (PFS) mounted on the 6.5m Magellan II Telescope at Las Campanas Observatory.

\subsection{3.9m University College London Échelle Spectrograph (UCLES)}

A total of 48 observations of HD208487 were collected using UCLES at the 3.9m Anglo-Australian Telescope (AAT) between JD 2451034.18 (August 8, 1998) to JD 2456969.96 (November 8, 2014) as part of the Anglo-Australian Planet Search (AAPS). UCLES is a cross-dispersed echelle spectrograph, which uses an $\text{I}_2$ absorption cell to derive RVs at the $\sim 3$m/s precision level \citep{AAPS}. In addition to bias subtraction and flatfielding, the $\text{I}_2$ method utilizes a point spread function (PSF) to properly wavelength calibrate the spectra. The known $\text{I}_2$ spectral features are used to calibrate the wavelengths of the stellar spectra and to model the PSF. Next, the stellar spectrum is deconvolved from the densely-packed $\text{I}_2$ features to successfully extract the correct Doppler shifts found in the observed stellar spectra, which enables the calculation of long-term stable, precise RVs \citep{IodineMethod}. UCLES offers a resolution of 40,000-120,000 depending on the slit width and CCD binning \citep{UCLES, TinneyDiscoveryHD208487}. The precision of UCLES data is between $2-3\text{ms}^{-1}$ \citep{Butler208487}. These data were obtained from the NASA Exoplanet Archive\footnote{https://exoplanetarchive.ipac.caltech.edu/} from \citep{Butler208487} and from newly measured RV data calculated for this work (Butler private communication).  These data will henceforth be referred to as UCLES. 

\subsection{3.6m High Accuracy Radial Velocity Planet Searcher (HARPS)}

HARPS is a stabilized, high-precision fiber-fed echelle spectrograph, contained in a vacuum cell (variations maintained $\leq 0.01$mbar) surrounded by a thermally stablized enclosure in order to avoid temperature fluctuations ($\leq 0.01$K) affecting the RV measurements \citep[see][]{HARPS_temp_press}. This instrument utilizes simultaneous observations through two fibers, one typically used to image the target and the other a Th-Ar reference spectrum to maintain high wavelength stability and monitoring over the long-term. With a fibre aperture on the sky of 1", HARPS maintains a resolving power of around 115,000. HARPS achieves an overall radial velocity accuracy of $\sim 1$m/s \citep{HARPS_original, HARPS_stability}. 

\subsubsection{TERRA}

Using publicly available data from HARPS obtained from the ESO HARPS archive (PIs: Mayor, M; Kuerster, M; Trifonov, T), we found 20 observations between JD 2453265.69 (September 17, 2004) to JD 2454258.94 (June 7, 2007). These observations were all processed with the built-in HARPS Data Reduction Software (DRS), which performs the standard bias, flat fielding, and wavelength calibration of the high-resolution spectra. The DRS also yields the extracted spectra, the cross-correlation functions (CCF), a measurement of the full-width at half maximum (FWHM), and the bisector inverse slope (BIS) \citep{HARPS_original, HD26965Diaz}. After the data passes through the DRS, radial velocity measurements are then re-computed using the HARPS-TERRA (Template-Enhanced Radial velocity Re-analysis Application) package \citep{TERRA}. Using all available orders of spectra of a star, HARPS-TERRA shifts them to zero, then co-adds the spectra, yielding one high S/N spectrum of the target. Next, the individual spectra are cross-correlated order by order against the high S/N stellar template to re-calculate the RVs.  We refer to these data as TERRA hereafter.

\subsubsection{TERRA2}

On June 3rd, 2015, a new fiber link was installed in HARPS in order to increase the stability of the instrument and the radial velocity precision  \citep{HARPS_upgrade}. As a result, we treat the observations from HARPS post-upgrade as a distinct instrument and refer to these radial velocity measurements as TERRA2, since they were processed in a similar manner to the pre-upgrade HARPS data. After the upgrade, we have a total of 106 observations spanning from JD 2457559.81 (June 20, 2016) to JD 2458755.48 (September 28, 2019).

\subsection{6.5m Carnegie Planet Finder Spectrograph (PFS)}

PFS is a high resolution, optical echelle spectrograph, which is outfitted with an $\text{I}_2$ cell. This Iodine absorption cell aids with wavelength calibrations to ensure precise radial velocity measurements are derived from the Doppler shift of certain spectral lines \citep{IodineMethod}. PFS is temperature controlled at $27 \pm 0.01^{\circ}$C and operated at ambient atmospheric pressure. The data taken prior to 2018 has a resolution of $\sim 80,000$ ($0.5"$ slit). In early 2018, the old 4K x 4K CCD with 15 micron pixels was replaced with a modernized 10K x 10K chip with 9 micron pixels. Since the upgrade\footnote{https://users.obs.carnegiescience.edu/crane/pfs/specs.php}, observations have been taken with a 0.3" slit, yielding a resolution of $\sim 130,000$ with a radial velocity uncertainty on the order of $\sim 1$ m\slash s \citep{PFS, Crane2008, PFS_new}, for targets brighter than V=10. In this study, we have 45 observations from JD 2455847.60 (October 13, 2011) until JD 2459508.64 (October 21, 2021). 

\section{Stellar properties}

\begin{table}
\centering
    \caption{Stellar parameters of HD208487}
    \begin{tabular}{lcc}
    \hline \hline Parameter & Value & Source \\
    \hline R.A. (J2000) & $21:57:19.85$ & SIMBAD \footnote{} \\
    Dec. (J2000) & $-37:45:49.05$ & SIMBAD \\
    $m_V$ & 7.47 & SIMBAD \\
    $B-V$ & 0.82 & SIMBAD \\
    Distance (pc) & $44.88$ & GAIA \footnote{} \\
    \hline Spectral type & G2V & \cite{TinneyDiscoveryHD208487} \\
    Mass $\left(M_{\odot}\right)$ & $1.15 \pm 0.04$ & This work (\ariadne) \\
    Radius $\left(R_{\odot}\right)$ & $1.16 \pm 0.02$ & This work (\ariadne) \\
    Age $(\mathrm{Gyr})$ & $1.95	\genfrac{}{}{0pt}{}{+1.41}{-1.11}$ & This work (\ariadne) \\
    Luminosity $\left(L_{\odot}\right)$ & $1.71 \pm 0.09$ & This work (\ariadne) \\
    $\mathrm{T}_{\text {eff }}(\mathrm{K})$ & $6134\genfrac{}{}{0pt}{}{+54}{-50}$ & This work (\ariadne) \\    
    $[\text{Fe} / \text{H}]$ & $0.09 \pm 0.04$ & This work (\ariadne) \\
    $\log g$ & $4.34\genfrac{}{}{0pt}{}{+0.62}{-0.42}$ & This work (\ariadne) \\
    $v \sin i\left(\mathrm{km}  \mathrm{s}^{-1}\right)$ & $5.58 \pm 0.27$ & This work (\species) \\
    $\log R_{H K}^{\prime}$ & -4.90 & \cite{HD208487_logrhk} \\
    \hline
    
    \end{tabular}
    
    \label{Stellar Param}
\end{table}
\footnotetext[4]{http://simbad.u-strasbg.fr/simbad/}
\footnotetext[5]{{This work has made use of data from the European Space Agency (ESA) mission
{\it Gaia} (\url{https://www.cosmos.esa.int/gaia}), processed by the {\it Gaia}
Data Processing and Analysis Consortium (DPAC,
\url{https://www.cosmos.esa.int/web/gaia/dpac/consortium}). Funding for the DPAC
has been provided by national institutions, in particular the institutions
participating in the {\it Gaia} Multilateral Agreement.}}

We derived the host star properties using two independent methods based on equivalent width measurements implemented in \species\footnote{https://github.com/msotov/SPECIES} \citep{SPECIES} and spectral energy distribution (SED) fitting with \ariadne\footnote{https://github.com/jvines/astroARIADNE} (Vines and Jenkins 2022).

\subsection{\species\ – Spectroscopic Parameters and atmosphEric ChemIstriEs of Stars}
\species\ computes atmospheric parameters (T$_{\rm eff}$, [Fe/H], log g, $\xi_{\rm t}$) as well as 11 elemental abundances from high resolution spectra \citep{SPECIES, speciesII}. Initially, the radial velocity is measured by cross-correlation with a G2 binary mask, and the spectrum is then shifted to the laboratory reference frame, where the Fe I and Fe II lines are identified and equivalent widths measured by fitting Gaussian-like profiles to the absorption lines. A grid of models taken from interpolating ATLAS9 \citep{ATLAS9} model atmospheres, as well as the equivalent widths, are provided to MOOG \citep{Sneden1973} in order to solve the radiative transfer equation (RTE) and estimate the correlation between Fe line abundances as a function of excitation potential and equivalent widths under the Local Thermodynamic Equilibrium (LTE) assumption. The above process is iterative, and is executed with a new set of parameters (T$_{\rm eff}$, [Fe/H], log g, $\xi_{\rm t}$) until either the number of maximum iterations is reached or when no correlation is detected above an established threshold $\epsilon$, where convergence is assumed for the corresponding stellar parameters. \species\ mass, radius and age are estimated using the Isochrones package \citep{Morton_isochrones} by interpolating through a grid of MIST \citep{MESA/MIST} evolutionary tracks using T$_{\rm eff}$, [Fe/H], log g as well as astrometric and photometric information. Finally, the v$\sin{i}$ and macro turbulent velocities are obtained from temperature calibrators and fitting Fe absorption lines of the observed HD208487 spectra with synthetic line profiles. \\
\\ HD208487 stellar parameters were derived from the 10 highest signal-to-noise (SN) spectra (SN$\sim200$) fetched from the ESO database\footnote{http://archive.eso.org/cms.html}, where \species\ outputs from every run being within the 1-$\sigma$ statistical agreement for the stellar parameters were recorded.  The typical median and uncertainties for the main stellar parameters from \species\ were T$_{\rm eff}$ = $6163 \pm 50$ K, [Fe/H] = $0.09 \pm 0.05$ dex, log g = $4.39  \pm 0.07 $, vsini = $5.63 \pm 0.28$ km s$^{-1}$, Age = $1.22^{0.61}_{-0.55}$ Gyr, M$_\star$ = $1.17 \pm 0.02$ M$_\odot$, and R$_\star$ = $1.15 \pm 0.01$ R$_\odot$.

\subsection{\ariadne\ – spectrAl eneRgy dIstribution bAyesian moDel averagiNg fittEr}
Adjusting the SED to a stellar atmosphere model is a common method to derive stellar properties. However, the physical assumptions of the underlying atmosphere model may contribute to bias the stellar parameters. Here we used a Bayesian Model Averaging approach implemented in \ariadne\ \citep{ARIADNE}, which fits archival photometry to several stellar atmosphere models, in order to mitigate the bias from a single atmosphere model. The stellar parameters were the results of an averaged posterior distributions from the Phoenix V2 \citep{Pheonix}, BT-Settl \citep{NextGen, Allard2012,ATLAS9_1993, ATLAS9} SED models weighted by their respective Bayesian evidence estimates. \\
\\ To derive the parameters from Table \ref{Stellar Param}, \species\ T$_{\rm eff}$, log g, [Fe/H] were used as priors along with archival photometric data, which was automatically fetched by \ariadne\ for the SED fitting. \ariadne\ outputs are used to derive the stellar age, mass and the equal evolutionary points from the isochrone package.

\section{Periodogram analysis}

    To begin the analysis of the RV data, we utilized the \exo\ package, which is a "transit and radial velocity interactive fitting tool for orbital analysis and N-body simulations" built as an accessible GUI \citep{exostriker}. We searched for embedded periodic signals in the data using the Generalized Lomb-Scargle (GLS) periodogram with a minimum period of 1 day and a maximum period of 10,000 days \citep{Lomb, Scargle, GeneralizedLS}.
\subsection{Full RV data set}
\label{Full RV GLS}
\begin{figure}
\includegraphics[width=\columnwidth]{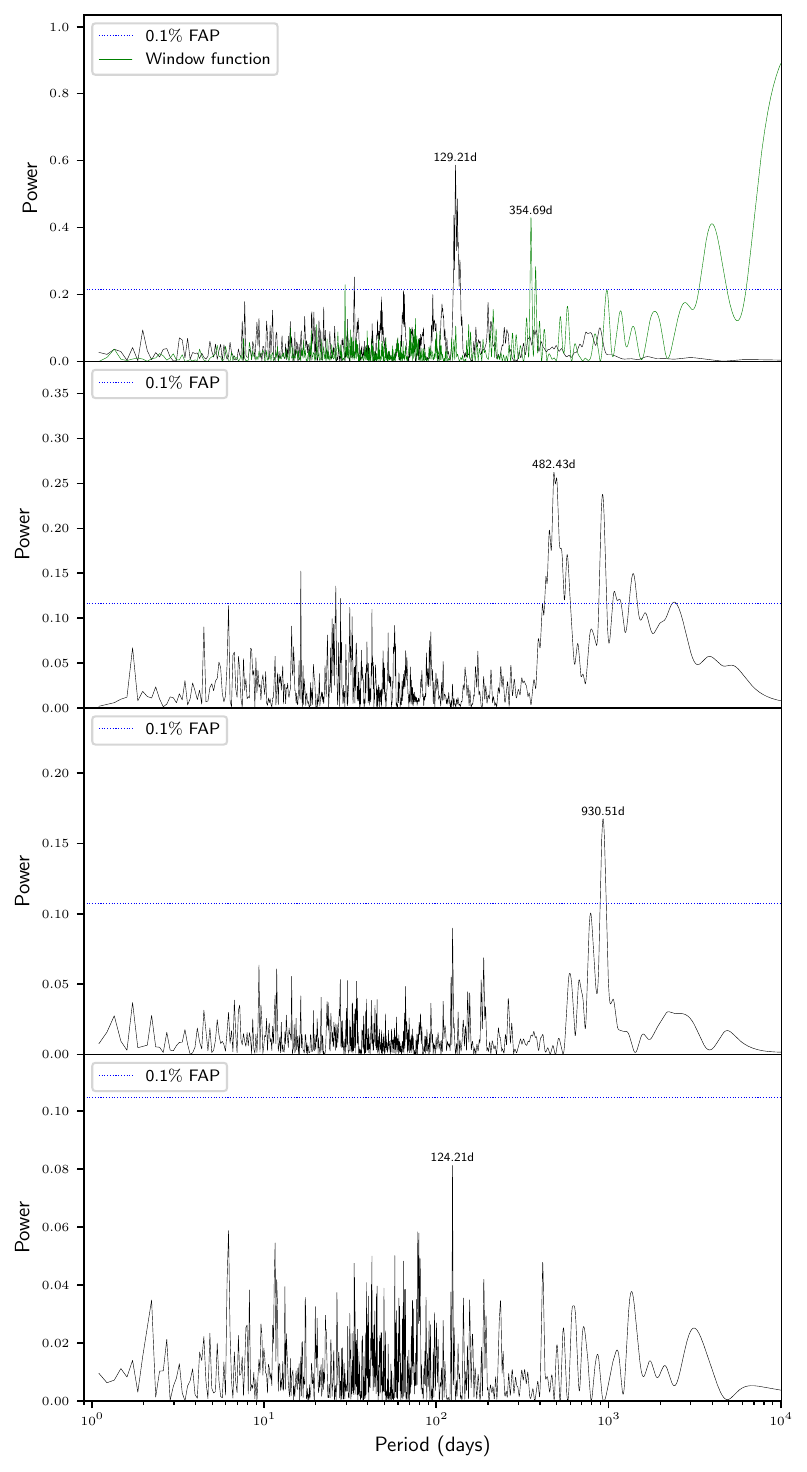} \\
\vspace{0.00mm}
\caption{Top panel: original GLS periodogram of the combined and mean-subtracted RV data (black). The periodogram of the window function (sampling) for the combined data (green) is also shown. Second panel: GLS periodogram after fitting and removing the first detected signal. Third panel: GLS periodogram after fitting and removing both the first and second signals. Bottom panel: GLS periodogram after fitting and removing all three previously detected signals. The blue, dotted horizontal lines represents the 0.1\% false alarm probability level estimated from 5000 bootstrap resamplings.}
\label{Periodograms}
\end{figure}

After inserting the combined RV data from UCLES, HARPS, and PFS into \exo, we mean-subtracted the RVs and then fit a linear trend to the data. The top panel in Figure \ref{Periodograms} shows the results of the original GLS periodogram, after the initial mean-subtraction and linear trend fit (with the Window function in green). The next panel displays the GLS periodogram of the residuals after fitting the first Keplerian (k1) signal, where the result shows the remaining embedded signals. Similarly, the third panel illustrates the GLS periodogram on the residuals of the combined fit of the k1 and k2 signals. Additionally, in that panel, we see flanking peaks that surround the $\sim 930$d signal. Although they appear to be statistically significant peaks, they are artifacts of the main signal due the discrete sampling of the data (window function) beating with the main signal. The bottom panel reveals that there are no remaining significant peaks after fully fitting a three-planet model. \\
\\
After fitting all the signals that surpass the 0.1\% FAP threshold, we find periods at $P_1 = 129.3$d, $P_2 = 493.2$d, and $P_3 = 939.7$d. This preliminary analysis helps inform what signals may appear during the more rigorous Bayesian analysis. 

\subsection{Split RV data set}
\label{S-index GLS}
As a first check to test whether stellar activity is adding noise to the RV data, or indeed causing the signals we are seeing, we repeated the periodogram analysis with split data sets, where the split is made after calculating the median S-index value in each data set (excluding UCLES data, which do not have S-index values recorded).  We then split the RVs in half: one data set with all RVs corresponding to S-index values less than the median and one data set with all RVs corresponding to S-index values greater than or equal to the median. We found that the low S-index data find all three signals, while the high S-index data only find the inner gas giant. Thus, the low S-index data provide a significantly better fit to the data when compared to the high S-index data, which would be expected if the RV signals are all independent of the stellar activity. 

\subsubsection{Low S-index}

The low S-index data set includes 10 TERRA, 22 PFS, and 53 TERRA2 observations, for a total of 85 RVs. Duplicating the analysis from the full data set, the GLS periodogram is used to search for any periodicity in this data set of lower stellar activity. To evaluate the models that we fit, we check the Bayesian Information Criteria (BIC). In general, a $\Delta BIC \geq 5$ ($\sim$150x more probable) is considered statistically significant \citep{BayesFactor, FengBIC}. After mean-subtracting the RVs and fitting a linear trend, the initial k0 model yields a BIC of 5940.85. The GLS periodogram clearly shows the first two signals above the 0.1\% FAP threshold, with the third slightly below, as seen in Figure \ref{LowS_Periodograms}. Fitting the first signal, with a period of $P_1 = 129.3d$, yields a BIC of 620.87 and a $\chi^2_{red} = 4.60$. Next, we add the second signal of $P_2 = 536.6d$ to the model, which after fully fitting ends with a period of $P_2 = 526.4d$, with a BIC of 452.98 and a $\chi^2_{red} = 2.25$. Initially, the $P_3$ signal falls below the 0.1\% FAP threshold. However, fitting the third period found from the highest peak and refitting all of the signals ends up reducing the BIC to 444.66, where $P_3 = 937.6d$. Adding the third signal reduced the BIC by 8.32 ($\sim$4100x improvement), enabling us to conclude that the third signal still exists in this data set. After fully fitting the three-planet model, we found that $rms = 2.18 m/s$, $\chi^2 = 137.94$, and $\chi^2_{red} = 2.03$, showing a good fit to the data.

\subsubsection{High S-index}

The high S-index data set includes 10 TERRA, 23 PFS, and 53 TERRA2 observations, for a total of 86 RVs.  The RVs corresponding to the data with higher stellar activity were examined with the GLS periodogram, finding that only the first of the previously detected peaks shows up at high statistical significance (Figure \ref{HighS_Periodograms}). A BIC of 3719.79 was found after mean-subtracting and fitting a linear trend to the RVs. Once again, we attempted to manually fit the peaks from the periods found with the full data set. After adopting the first period of $P_1 = 130.3$d as the k1 signal, the best fit is found at $P_1 = 130.6d$ with a BIC of 1020.59 and a $\chi^2_{red} = 9.67$, showing the signal is significant within the BIC test. Neither of the remaining two signals are detected from the high S-index data. Thus, we concluded that the stellar activity is actually obscuring the signals, rather than contributing to them.

\section{Bayesian analysis}

\begin{figure*}
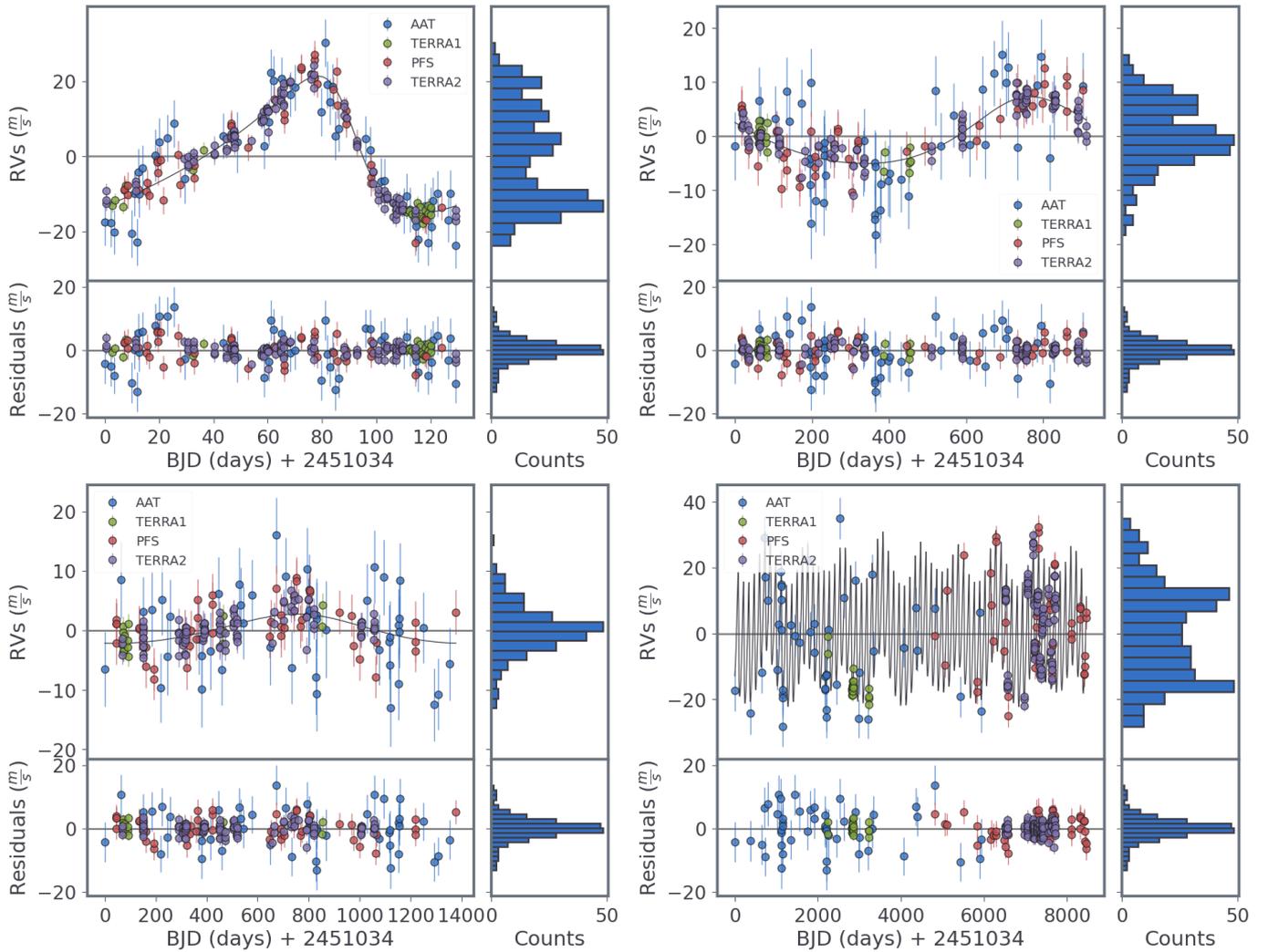


\includegraphics[width=\columnwidth]{rv_phased_K1.pdf} 
\includegraphics[width=\columnwidth]{rv_phased_K2.pdf} \\
\includegraphics[width=\columnwidth]{rv_phased_K3.pdf} 
\includegraphics[width=\columnwidth]{rv_timeseries.pdf}
\caption{Top to bottom, left to right: phase-folded, best-fit model for the first, second, third, and complete Keplerian model.}
\label{Emperor Curve Fit}
\end{figure*}
\begin{figure*}

\includegraphics[width=0.7\columnwidth]{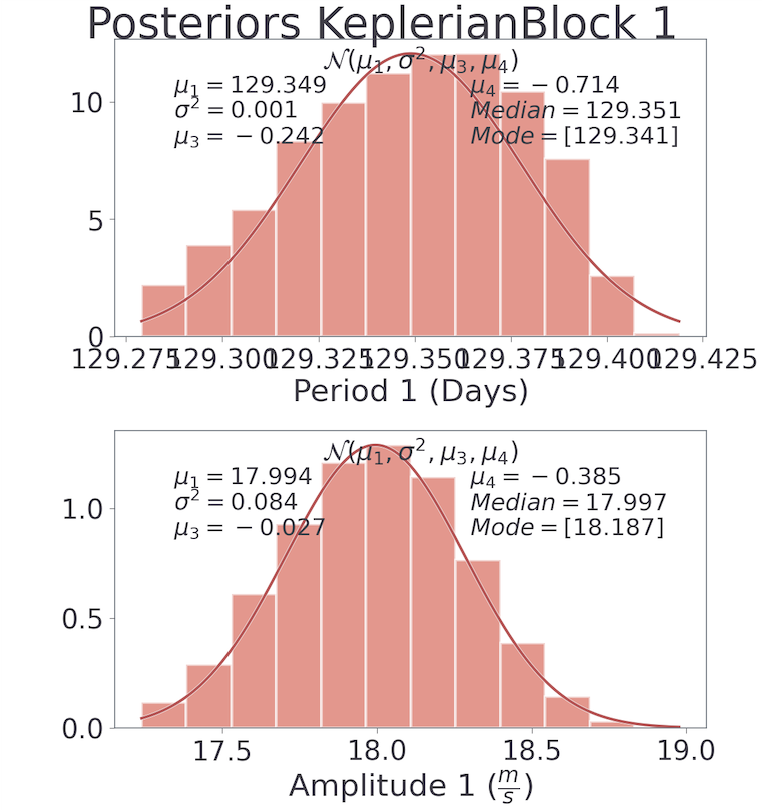} 
\includegraphics[width=0.7\columnwidth]{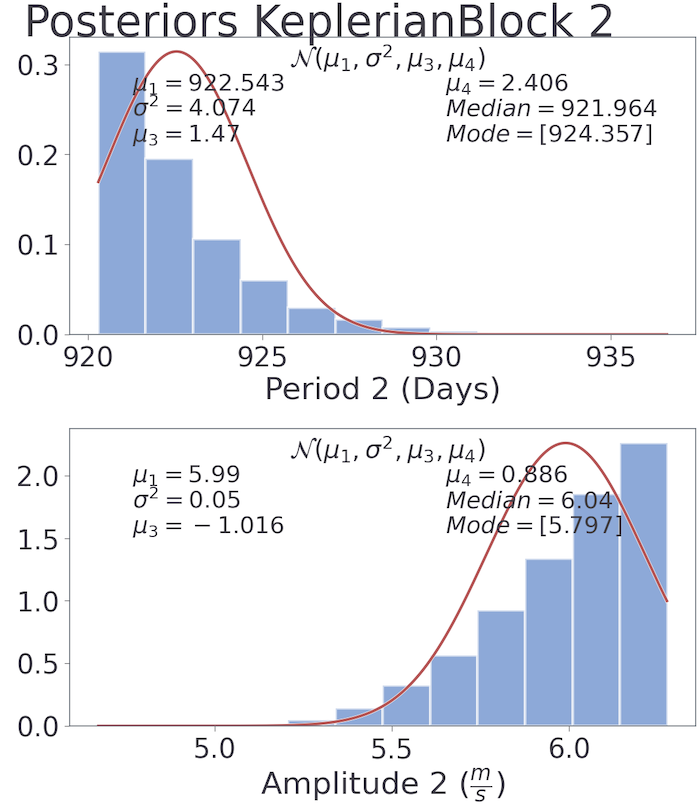} 
\includegraphics[width=0.7\columnwidth]{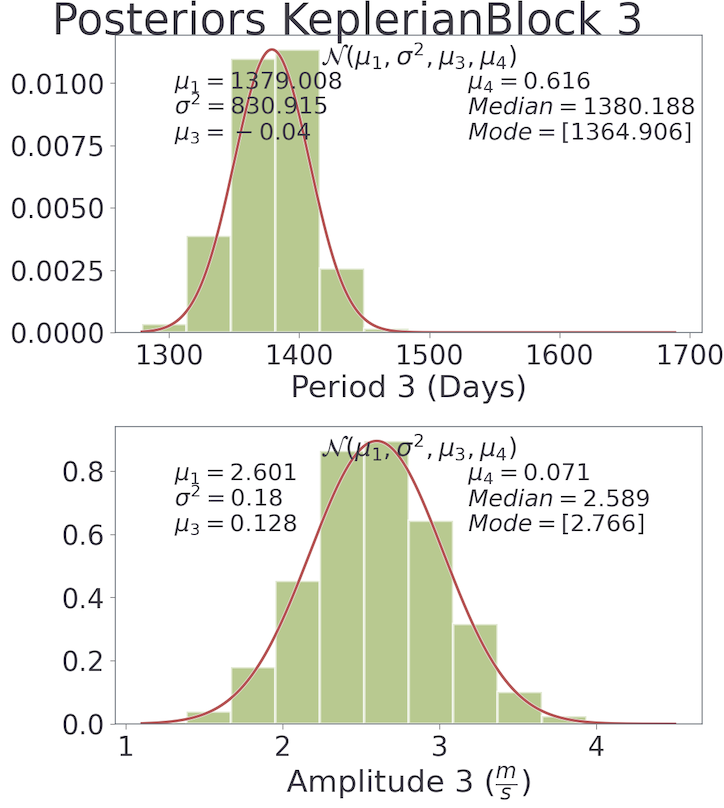} \\

\caption{Top, left to right: final posterior distribution of the period for the first, second, and third Keplerians from the best-fitting parallel-tempered MCMC run. Bottom, left to right: final posterior distribution of the semi-amplitude for the first, second, and third Keplerians from the best-fitting parallel-tempered MCMC run. The numbers at the top of each figure correspond to the mean, variance, skewness and kurtosis, median, and mode, respectively. The solid line represents a Gaussian curve with the same mean and variance.}
\label{Posteriors}
\end{figure*}
After examining the results from the periodogram analyses, we found significant evidence for two additional planetary signals in the data, along with the previously confirmed planet, HD208487b. In order to corroborate the initial results, we followed similar procedures to those employed in \cite{Tuomi_Bayesian, jenkins2014, MagellanPFS_Diaz} to model the RVs. The Markov Chain Monte Carlo (MCMC) code, Exoplanet Mcmc Parallel tEmpering Radial velOcity fitteR\footnote{https://github.com/ReddTea/astroemperor/tree/master} \citep{pena2025}, "searches for signals in radial velocity timeseries, employing Adaptive Parallel Tempering MCMC, convergence tests and Bayesian statistics, along with various noise models." Further, the Python code provides Bayesian posterior distribution models and the parameterization for the planetary signals to describe the best-fit orbital solution to the star system. An early version of \emp\ has already been used in a number of works to detect small signals in RV data \citep{Wittenmyer_astroemperor,proxcentb_emp2, GJ357_emp3, DMPP3_emp4, LTT_emp5}, where the updated version employs a custom-written Adaptive Parallel Tempering MCMC engine, similar to the \emcee\ package \citep{emcee}, as used in \citep{Gregory_longerP, Gregoryupdated}. \\
\\
We successively ran the \emp\ code with different properties to find the optimal setup. The statistical models included a function describing a $k$-Keplerian planet model, a linear trend term (to account for any existing accelerations within the system), a red-noise model (consisting of a moving average model with exponential weighted smoothing), and linear correlations with the stellar activity indices. The full mathematical description of the model with all parts included is shown below: 
\begin{align}  \label{eq:full_model}
    y_{i, ins}&=\gamma_{ins}+\dot{\gamma} t_i+f_k(t_i)+\sum_{n=1}^{q_{ins}} c_{n, ins} \xi_{n, i, ins} \nonumber \\
&+\sum_{l=1}^p \phi_{ins, l} \exp \{\frac{t_{i-l}-t_i}{\tau_{ins}}\} r_{i-l, ins}+\epsilon_{i, ins},
\end{align}
where $y_{i,ins}$ corresponds to the observation at time $t_i$ for each of the distinct instruments, $\gamma_{ins}$ is the velocity offset for each of the instruments, $\dot{\gamma}$ is a linear trend term (for any existing acceleration), and $c_{n,ins}$ describes the linear correlations with $q$ stellar activity indices $\xi_{n,i,ins}$ for each instrument. $r_{i,ins}$ denotes the residuals once the model has been subtracted from the measurement. The function $f_k$ is a superposition of $k$-Keplerian signals denoted by:
\begin{equation}
f_k\left(t_i\right)=\sum_{m=1}^k K_m\left[\cos \left(\omega_m+\nu_m\left(t_i\right)\right)+e_m \cos \left(\omega_m\right)\right]
\end{equation}
where $K_m$ is the velocity semi-amplitude, $\omega_m$ is the longitude of periastron, $\nu_m$ is the true anomaly and $e_m$ is the eccentricity. Since $\nu_m$ is a function of the orbital period and the mean anomaly $M_{0,m}$ at epoch, we can fully describe $f_k$ by $K_m$, $\omega_m$, $e_m$, $M_{0,m}$, $P_m$; $m \in \{1,...,k\}$. \\
\\
The white noise term is denoted by the additive random variable $\epsilon_{i,ins}$, shown by:
\begin{equation}  \label{eq:wno_model}
    \epsilon_{i,ins} \sim \mathcal{N}(0,\sigma_i^2+\sigma_{ins}^2)
\end{equation}
where $\sigma_i$ is the uncertainty reported with the measurement $y_{i,ins}$ and $\sigma_{ins}$ is the excess white noise or jitter modelled after each instrument. This nuisance parameter, which is required to model the mechanisms that create the data, has a Normal prior bound (Table \ref{Priors}), in the model. The remaining terms define the rest of the noise model, as described in \cite{HD26965Diaz}.

\subsection{Posterior samplings and signal detection}

\begin{table}
    \caption{Prior selection for the parameters}
    \begin{tabular}{lcc}
    
    \hline \hline Parameter & Prior Type & Range \\
    \hline Semi-amplitude & Uniform & $K \in\left(0, 100\right]$ \\
    Period & Uniform & $P \in\left[1,8475\right]$ \\
    Eccentricity & $\mathcal{N}\sim\left(0, \sigma_e\right)$ & $e \in[0,1)$ \\
    Long. of Peric. & Uniform & $\omega \in[0,2 \pi]$ \\
    Mean Anomaly & Uniform & $M_0 \in[0,2 \pi]$ \\
    Jitter & $\mathcal{N}\sim\left(5, 5\right)$ & $\sigma_{ins} \in\left(0, 100\right]$ \\
    Offset & Uniform & $\gamma_{ins}\in [-100, 100]$ \\
    Acceleration & Uniform & $a \in [-1,1]$\\
    Stellar Activity & Uniform & $SA \in [-1,1]$\\
    \hline
    
    \end{tabular}    
    \label{Priors}
\end{table}

The Adaptive Parallel Tempering provides a ladder of replicas of the system, where each step has the likelihood at a different temperature. The hotter the chain, the flatter the posterior, allowing the sampler to jump from one local maxima to another, whilst the cold chains explore the phase space in detail. Swapping states for random members of different systems makes the cold chain fully characterize each of the maxima. The adaptive scheme utilizes varying temperatures to serve as a 'good ladder.' This ladder allows for the calculation of the Bayesian evidence via thermodynamic integration as well (more details in \cite{Vousden}). The prior distributions for the orbital and instrumental parameters for our model are listed in Table $\ref{Priors}$.

\subsection{Model selection}
\label{Full RV emp}
The final piece of the \emp\ code deals with choosing the proper model. Ideally, the final model represents the most accurate description of the orbital and planetary parameters. As a result, \emp\ performs model selection at each stage to ensure the current model actually produces an improvement over the previous model. In order for the new $k+1$ model to be preferred over the existing $k$ model, we require an improvement over a statistic of choice. In this scenario, for the model comparison, the log of the evidence (Z) was chosen, and a Bayes factor of at least 5 was required, which represents a $150\times$ more probable solution than the previous model. The model evidence, or marginalized likelihood, corresponds to the likelihood of generating the data from the prior, given a correct model. In other words, it could be understood as the probability of the model itself, hence its use for model comparison \citep{robert_2007}. The ratio of evidences for two competing models is called the Bayes factor and is used for model comparison.\\
\\
The \emp\ MCMC code employs a burn-in stage, in which the chains search for the high-probability zones, whilst adapting their temperatures to have an efficient swap rate. When the burn-in stage is over, the samples are discarded, and temperatures stop adapting. This ensures that the chain remains ergodic whilst satisfying detailed balance. This process lasts for 80\% of the total iterations. In our analysis, the Markov-chains comprised 12 temperatures, 1200 walkers, and 20000 steps, combining for 288 million iterations per model (with a 230 million iteration burn-in). Then, the chain must be longer than 50 times the autocorrelation time, or it is extended. The models we ran, beyond the straight Keplerian models include:
\begin{enumerate}
    \item a white-noise model (WNO, as shown in eq. \ref{eq:wno_model})
    \item correlated noise, modelled by an exponentially weighted moving average (MAO, see last term of eq.\ref{eq:full_model})
    \item white noise with linear stellar activity correlations (CorrO, see the first sum of eq.\ref{eq:full_model})
    \item correlated noise plus the linear stellar activity correlations (CorrMA)
\end{enumerate} 
 \textcolor{white}{.} \\
 \\
Initially, the Corr models were run with every available stellar index, however only the indices with correlations with the RVs that were inconsistent with zero (no matter how low the rank) were included in the final set of runs. As a result, the Corr models are based on the S-index from PFS and the BIS from TERRA1 and TERRA2. \\
\\
Additionally, we ran each of these models with a fixed eccentricity of zero, a tightly constrained eccentricity described by a normal distribution centered at zero with a standard deviation of 0.1, and a constrained eccentricity described by a normal distribution centered at zero with a standard deviation of 0.3 (as explained in \cite{jenkins2025}), which enables \emp\ to explore a wide variety of solutions based on the potential planetary eccentricities. In general, we found that fixing the eccentricity prior to zero would cause additional signals to appear that were harmonics of the real signals, primarily a consequence of the fact that the largest signal in the data corresponds to a massive planet on a non-circular orbit, and hence is poorly explained by a strict cosine-like function. All circular models that caused the 1st harmonic of this primary signal (HD208487b) to appear as a secondary signal in the data performed statistically much worse, enabling us to securely disregard these harmonic models. On the other hand, most of the models with the constrained eccentricity prior or the tightly constrained eccentricity prior yield very similar results. The main difference between the constrained and tightly constrained models are the predicted eccentricities of the outer two planets. The periods and semi-amplitudes are similar, but the constrained models tend to terminate with higher eccentricities, which results in both worse final statistics and long-term unstable planetary dynamics. In the theoretical limit of infinite walkers and steps, we would expect all of the models to converge to the same parameter configuration, even though we extended our analyses as far as reasonably possible to 288 million iterations. \\
 \\
 We found that the CorrO with the tightly constrained eccentricity model performed the best, yielding a dynamically stable configuration.
 An independent verification method was performed to validate our results, which was in agreement with our best-fit model (see \S \ref{Independent verification}). The final orbital and instrumental parameters for the best-fitting model (CorrO-tightly constrained eccentricity) are shown in Table \ref{Final Params}. The best-fit Keplerians are shown in Figure \ref{Emperor Curve Fit} and the posterior distributions for the periods and semi-amplitudes are shown in Figure \ref{Posteriors}. 

\subsubsection{Instrument contributions to the model}
Due to the varying number of RVs in each dataset as well as the differences in quality of each instrument, it is worthwhile to determine each dataset's overall contribution to the final model based on each instrument's RMS. Each dataset's contribution is approximated by dividing its own number of observations by the square of its RMS, ($\frac{N_i}{RMS_i^2}$) and then finding the weighted sum. As shown in Table \ref{Contributions}, although AAT and PFS comprise around one fifth of the overall number of RVs, they both contribute a disproportionately low amount to the overall best-fit model. In fact, the AAT data contributes $7\times$ less value than expected based on the number of RVs. On the other side, the TERRA2 data contributes most to the model due to its high precision, and large data volume. However on the other hand, including the AAT data increases the duration of the baseline from about three to five full periods of the outer planet, which lowers the uncertainties of the orbital parameters, especially for the longer-period planets.  Therefore, we find an interesting balance between modeling long-baseline, yet lower precision RV datasets, against short-baseline but high-precision measurements, whereby even though some datasets can be an order of magnitude lower in precision, their overall observing baselines can overcome these short comings when targeting longer-period worlds.

\subsubsection{Comparison of earlier data and later data}
\label{Time split emp}
Checking the stability of the RV signals over time provides another important determination of the veracity of the signals. We split the complete RV data set into first half and second half data, separated by instrument. Thus, the first half data includes 68 RVs, 48 data points from AAT and 20 data points from TERRA. The second half data includes 151 data points, 45 RVs from PFS and 106 RVs from TERRA2. We executed the \emp\ code on the split data sets, testing the various types of models previously discussed. In the first half data, the best fit model finds three statistically significant signals, $P_1 = 129.39\genfrac{}{}{0pt}{}{+0.15}{-0.22}$d
, $P_2 = 934.56 \genfrac{}{}{0pt}{}{+17.97}{-42.29}$d
, and $P_3 = 1294.65 \genfrac{}{}{0pt}{}{+1191.79}{-38.62}$d 
(see Table \ref{Final Params}).\\
\\
The second half data, reveals similarities to the orbital parameters derived from the first half data set. The periods and semi-amplitudes of the fit to the second-half data are as follows: $P_1 = 129.34\genfrac{}{}{0pt}{}{+0.01}{-0.03}$d
, $P_2 = 485.60 \genfrac{}{}{0pt}{}{+4.71}{-6.68}$d
, and $P_3 = 979.74 \genfrac{}{}{0pt}{}{+33.79}{-36.32}$d 
(see Table \ref{Final Params}). The period and semi-amplitude of HD208487b remain almost identical between the early and late data. The period and semi-amplitude of HD208487c are also consistent within uncertainties. The main difference between the two split data sets is the appearance of HD208487d as an intermediate-period signal rather than a long-period signal. Interestingly, the later data set, with a baseline of only 3661d, is in close agreement with the initial \emp\ model, which performs worse statistically than the chosen model. However, the model based on the earlier data, with a longer baseline of 5936d, more closely resembles the chosen CorrO model, as evidenced by the similar periods for all three signals.  We note that since this intermediate signal repeatedly appears, just with lower statistical significance, this may indicate the reality of a further Doppler signal that requires more high-precision RVs to confirm.

\begin{table*}
\centering
\caption{Comparison of the best WNO models (with $1 \sigma$ errors) for different dataset aggregations. The overall best fit of orbital and instrumental parameters solutions for HD208487 is listed under ALL.}
\begin{tabular}{lcccccc}
\toprule
Parameter           &TERRA+TERRA2            &No AAT                  &AAT+TERRA               &PFS+TERRA2              & AAT only                      & ALL\\
\midrule
\midrule

$P$ (days)          &$129.33^{+0.04}_{-0.04}$&$129.32^{+0.03}_{-0.01}$&$129.39^{+0.15}_{-0.22}$&$129.34^{+0.01}_{-0.03}$&$129.27^{+0.23}_{-0.14}$&$ 129.36^{+0.02}_{-0.02}$\\
$K$ (m/s)           &$18.38^{+0.08}_{-0.23}$ &$18.76^{+0.27}_{-0.35}$ &$17.35^{+0.95}_{-1.22}$ &$18.53^{+0.10}_{-0.17}$ &$16.70^{+0.95}_{-0.02}$ &$18.28^{+0.29}_{-0.40}$ \\
$e$                 &$0.36^{+0.01}_{-0.02}$  &$0.35^{+0.01}_{-0.01}$  &$0.18^{+0.04}_{-0.06}$  &$0.35^{+0.00}_{-0.01}$  &$0.13^{+0.03}_{-0.02}$  &$0.37^{+0.01}_{-0.01}$  \\
$\omega$ (rad)      &$1.21^{+0.03}_{-0.01}$  &$1.21^{+0.01}_{-0.03}$  &$1.41^{+0.55}_{-0.35}$  &$1.16^{+0.06}_{-0.04}$  &$2.58^{+0.48}_{-0.65}$  &$1.10^{+0.01}_{-0.01}$  \\
$M_0$ (rad)         &$3.43^{+0.08}_{-0.10}$  &$3.40^{+0.07}_{-0.01}$  &$1.83^{+0.38}_{-0.68}$  &$3.23^{+0.04}_{-0.05}$  &$0.58^{+0.78}_{-0.49}$  &$1.95^{+0.05}_{-0.04}$  \\
$m \sin i$ ($M_J$)  &$0.47^{+0.01}_{-0.01}$  &$0.48^{+0.01}_{-0.01}$  &$0.47^{+0.03}_{-0.03}$  &$0.47^{+0.01}_{-0.01}$  &$0.45^{+0.03}_{-0.00}$  &$0.46^{+0.01}_{-0.00}$  \\
$AU1$               &$0.52^{+0.00}_{-0.00}$  &$0.52^{+0.00}_{-0.00}$  &$0.53^{+0.00}_{-0.00}$  &$0.53^{+0.00}_{-0.00}$  &$0.52^{+0.00}_{-0.00}$  &$0.53^{+0.00}_{-0.00}$  \\
\midrule
$P_2$ (days)        &$490.82^{+5.62}_{-0.31}$&$488.10^{+0.89}_{-1.54}$&$934.56^{+17.97}_{-42.29}$&$485.60^{+4.71}_{-6.68}$&$923.92^{+26.38}_{-19.30}$ &$923.06^{+2.02}_{-2.76}$\\
$K_2$ (m/s)         &$3.73^{+0.27}_{-0.01}$  &$4.35^{+0.33}_{-0.46}$  &$9.76^{+1.26}_{-1.98}$    &$4.03^{+0.17}_{-0.09}$  &$8.69^{+1.13}_{-0.21}$     &$6.18^{+0.09}_{-0.07}$  \\
$e_2$               &$0.17^{+0.02}_{-0.06}$  &$0.30^{+0.06}_{-0.10}$  &$0.17^{+0.06}_{-0.12}$    &$0.22^{+0.01}_{-0.04}$  &$0.08^{+0.03}_{-0.03}$     &$0.19^{+0.01}_{-0.02}$  \\
$\omega_2$ (rad)    &$4.55^{+0.01}_{-0.40}$  &$4.72^{+0.33}_{-0.48}$  &$3.89^{+1.20}_{-0.04}$    &$4.00^{+0.14}_{-0.17}$  &$2.94^{+1.46}_{-1.10}$     &$5.96^{+0.30}_{-5.96}$  \\
$M_{02}$ (rad)      &$0.74^{+0.90}_{-0.98}$  &$0.41^{+0.19}_{-0.14}$  &$3.61^{+1.34}_{-1.70}$    &$2.39^{+0.00}_{-0.32}$  &$3.96^{+0.00}_{-1.07}$     &$1.29^{+0.52}_{-0.29}$  \\
$m_2 \sin i$ ($M_J$)&$0.16^{+0.01}_{-0.01}$  &$0.18^{+0.01}_{-0.01}$  &$0.51^{+0.07}_{-0.07}$    &$0.17^{+0.01}_{-0.01}$  &$0.46^{+0.06}_{-0.01}$     &$0.32^{+0.01}_{-0.01}$  \\
$AU2$               &$1.28^{+0.01}_{-0.01}$  &$1.27^{+0.01}_{-0.01}$  &$1.96^{+0.03}_{-0.03}$    &$1.27^{+0.01}_{-0.01}$  &$1.95^{+0.04}_{-0.03}$     &$1.94^{+0.01}_{-0.01}$  \\
\midrule
$P_3$ (days)        &$959.17^{+419.38}_{-194.63}$&$941.38^{+28.27}_{-18.51}$&$1294.65^{+1191.79}_{-38.62}$&$979.74^{+33.79}_{-36.32}$&$186.57^{+1125.53}_{-186.57}$&$1380.13^{+19.20}_{-8.25}$\\
$K_3$ (m/s)         &$3.16^{+1.01}_{-0.94}$      &$3.54^{+0.36}_{-0.50}$    &$5.61^{+1.63}_{-3.77}$       &$3.09^{+0.42}_{-0.37}$    &$4.35^{+0.07}_{-1.57}$       &$2.46^{+0.30}_{-0.09}$  \\
$e_3$               &$0.03^{+0.06}_{-0.05}$      &$0.22^{+0.08}_{-0.17}$    &$0.06^{+0.03}_{-0.02}$       &$0.07^{+0.04}_{-0.03}$    &$0.01^{+0.08}_{-0.01}$       &$0.11^{+0.03}_{-0.06}$  \\
$\omega_3$ (rad)    &$1.08^{+2.25}_{-1.69}$      &$0.57^{+2.24}_{-1.65}$    &$2.80^{+1.49}_{-0.65}$       &$3.34^{+0.12}_{-1.12}$    &$4.91^{+1.86}_{-3.09}$       &$0.29^{+0.28}_{-0.09}$  \\
$M_{03}$ (rad)      &$3.30^{+0.80}_{-1.37}$      &$3.27^{+0.37}_{-0.82}$    &$5.47^{+1.85}_{-3.32}$       &$5.67^{+2.17}_{-4.10}$    &$3.79^{+0.37}_{-2.08}$       &$2.61^{+0.47}_{-0.42}$  \\
$m_3 \sin i$ ($M_J$)&$0.168^{+0.07}_{-0.07}$     &$0.18^{+0.02}_{-0.02}$    &$0.33^{+0.10}_{-0.10}$       &$0.16^{+0.02}_{-0.02}$    &$1.34^{+0.11}_{-0.10}$       &$0.15^{+0.01}_{-0.02}$  \\
$AU3$               &$1.995^{+0.25}_{-0.25}$     &$1.97^{+0.03}_{-0.03}$    &$2.436^{+1.52}_{-1.52}$      &$2.00^{+0.05}_{-0.05}$    &$0.67^{+1.78}_{-0.67}$       &$2.54^{+0.02}_{-0.01}$  \\
\midrule
$\gamma_{UCLES}$ (m/s) & -                         &-                       &$18.06^{+2.71}_{-1.83}$    &-                         &$21.28^{+0.06}_{-1.53}$&$1.78^{+0.37}_{-0.94}$   \\
$\gamma_{TERRA1}$ (m/s)& $1.18^{+2.07}_{-0.19}$    &$1.87^{+0.69}_{-0.16}$  &$31.28^{+1.79}_{-0.21}$    &-                         &-                      &$12.10^{+0.41}_{-0.76}$   \\
$\gamma_{TERRA2}$ (m/s)& $-18.20^{+28.31}_{-15.94}$&$-14.78^{+2.93}_{-2.50}$&-                          &$3.91^{+0.20}_{-0.91}$    &-                      &$-8.373^{+0.74}_{-1.99}$   \\
$\gamma_{PFS}$ (m/s)   & -                         &$-15.86^{+3.25}_{-2.49}$&-                          &$3.18^{+0.30}_{-0.88}$    &-                      &$-8.65^{+0.70}_{-2.41}$   \\
$\dot{\gamma}$ (m/s yr)& $0.00^{+0.00}_{-0.01}$    &$0.00^{+0.00}_{-0.00}$  &$0.00^{+0.00}_{-0.00}$     &$-0.00^{+0.00}_{-0.00}$   &$-0.00^{+0.00}_{-0.00}$&$0.00^{+0.00}_{-0.00}$    \\
$\sigma_{UCLES}$ (m/s) & -                         &-                       &$4.94^{+0.99}_{-0.70}$     &-                         &$4.41^{+0.92}_{-0.63}$ &$5.65^{+0.81}_{-0.67}$    \\
$\sigma_{TERRA1}$ (m/s)& $2.03^{+0.45}_{-0.38}$    &$1.93^{+0.50}_{-0.37}$  &$2.49^{+0.03}_{-0.36}$     &-                         &-                      &$1.61^{+0.18}_{-0.13}$    \\
$\sigma_{TERRA2}$ (m/s)& $1.33^{+0.09}_{-0.05}$    &$1.13^{+0.24}_{-0.14}$  &-                          &$1.25^{+0.14}_{-0.13}$    &-                      &$1.38^{+0.17}_{-0.15}$    \\
$\sigma_{PFS}$ (m/s)   & -                         &$3.15^{+0.58}_{-0.45}$  &-                          &$3.66^{+0.10}_{-0.31}$    &-                      &$3.12^{+0.29}_{-0.12}$    \\

\midrule
BIC                    & 595.540                   &850.161                 &469.948                   &758.665                    &358.478                &1209.520        \\
$\log{Z}$              & -288.352                  &-424.660                &-223.560                  &-369.246                   &-169.400               &-582.577        \\
RMS                    & 1.806                     &2.431                   &4.568                     &2.553                      &4.915                  &3.741           \\
$\chi^2_{red}$         & 1.117                     &1.203                   &1.188                     &1.103                      &1.467                  &1.172           \\
\bottomrule

\end{tabular}
\label{Final Params}
\end{table*}

\begin{table}
    \caption{Contribution of each instrument to the best-fit model.}
    \begin{tabular}{lcccc}

                      \hline \hline & AAT  & TERRA & PFS  & TERRA2 \\
\hline
RVs                    & 48   & 20    & 45   & 106    \\
RMS                    & 6.57 & 1.62  & 3.58 & 1.85   \\
RV dataset \% & 22   & 9     & 21   & 48     \\
Contribution \%        & 3    & 18     & 8    & 72  \\
\hline    
    \end{tabular}    
    \label{Contributions}
\end{table}

\subsubsection{AAT-only run}
\label{AAT-only}

Ensuring consistency between earlier research and the current analysis remains an essential aspect of follow-up studies. Initially, \cite{TinneyDiscoveryHD208487} detected HD208487b utilizing a Periodogram analysis. Later, \cite{Gregory_longerP, Gregoryupdated} claimed the detection of HD208487c with a more statistically robust MCMC method compared to Tinney's Periodogram analysis. To maintain consistency, we re-analyzed the AAT data alone with \emp. The results from these MCMC runs, summarized in Table \ref{Final Params}, indicate very similar findings, adding confidence to the past research.

\subsection{Independent verification of results}
\label{Independent verification}
In order to increase the robustness and reliability of the results, we analyzed the data with independent sampling methods and codes. We applied the delayed-rejection adaptive Metropolis algorithm \citep{haario2006} to detect maxima in the posterior probability density and the simple adaptive-Metropolis algorithm \citep{haario2001} to estimate parameters and obtain solutions \citep[see e.g.][]{butler17,tuomi18}.\\
\\
We also increased the set of available activity-indices by calculating the "differential velocities" of \citet{feng2017} for the TERRA reduction of HARPS. This is to say that we divided the 72 HARPS orders into three equal subsets, corresponding to the red-most, middle, and blue-most parts of the corresponding spectra. These were then used independently to obtain three sets of radial velocities denoted as $y_{i,1}, y_{i,2}$, and $y_{i, 3}$ for $i=1, ..., N_{\rm HARPS}$. The differential velocities are thus $\xi_{i,1} = y_{i,1} - y_{i,2}$ and $\xi_{i,2} = y_{i,2} - y_{i,3}$ that can be used as additional activity indices. Because any and all Doppler signals of planets have been subtracted when producing these indices, the remaining variability simply corresponds to noise and activity-induced variations. In particular, wavelength-dependent variations that are present in these indices enable one to filter them out from the data by accounting for the respective correlations with these indices in accordance with Eq. (\ref{eq:full_model}). The models are thus equivalent except for the fact that the additional activity indices enable removing some additional activity-induced variability. Such variability indeed was found to be only weakly present in the HARPS data. Yet, since the results were consistent, the postulated planetary signals presented in the manuscript show no evidence for variations as a function of spectral wavelength.\\
\\
The results from these analyses are consistent with those presented in Table \ref{Final Params}. The obtained solution for the three-Keplerian model contains periods of 129.40$\pm$0.04, 917.0$\pm$6.4, and 1396$\pm$27 days, respectively. These estimates are consistent with the results obtained with the \emp\ code, albeit slightly more uncertain. This excess uncertainty may be caused by the fact that the model contains additional activity indices and is thus more flexible. It is also clear that wavelength-dependent variations in the HARPS radial velocities, as estimated by the differential velocities, are not responsible for the observed Keplerian periodicities.

\section{Stellar activity and correlations}
\subsection{GLS periodograms and correlations of stellar activity}
\begin{figure}
\includegraphics[width=0.95\columnwidth]{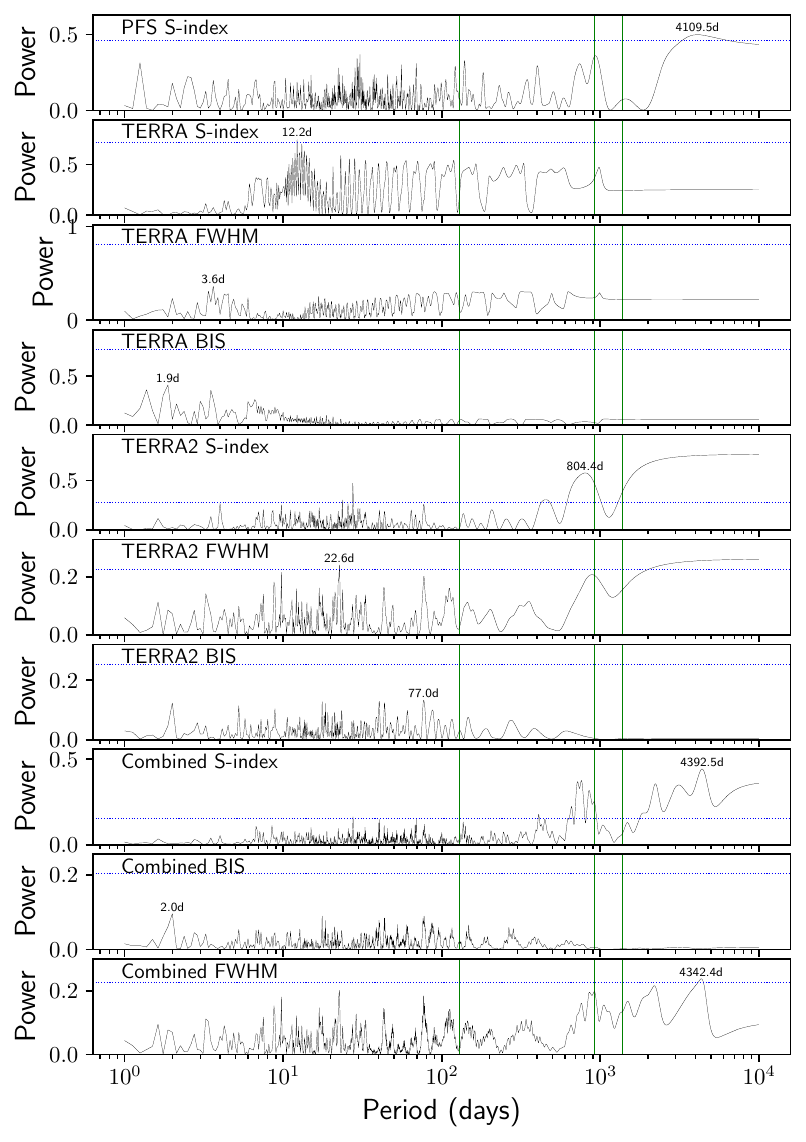}

\caption{GLS periodogram of the available stellar activity indices. From top to bottom: PFS S-index, TERRA S-index, TERRA FWHM, TERRA BIS, TERRA2 S-index, TERRA2 FWHM, TERRA2 BIS, Combined S-index, Combined BIS, Combined FWHM. The green, solid vertical lines show the best-fit periods of the three planetary signals. The blue, dotted horizontal lines represents the 0.1\% false alarm probability (FAP) level estimated from 5000
bootsrap resamplings.}
\label{Stellar Activity Periodograms}
\end{figure}

\begin{figure*}
\begin{center}
    \includegraphics[width=0.7\columnwidth]{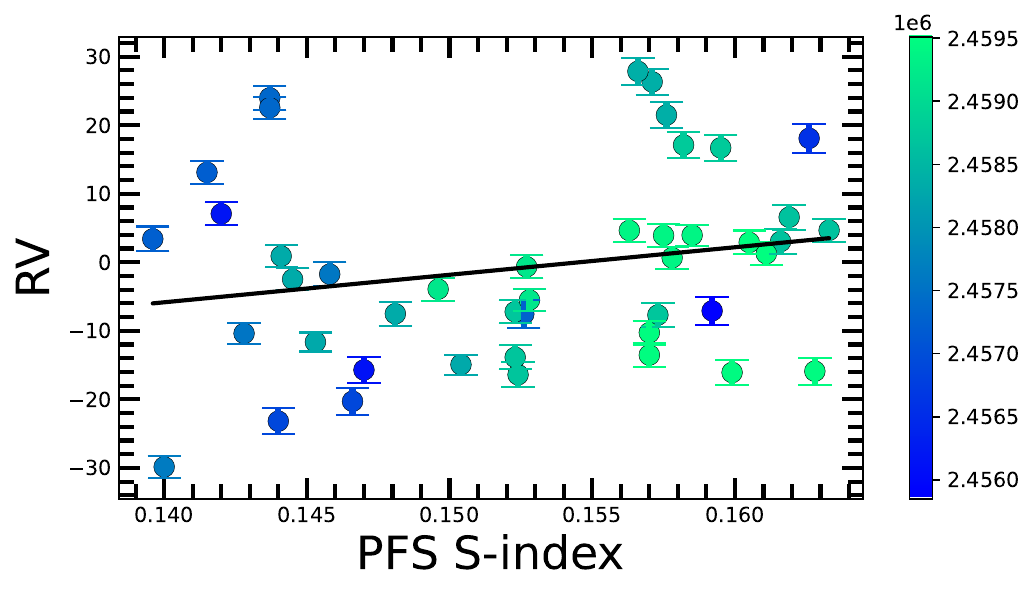}
\end{center} 
\includegraphics[width=0.7\columnwidth]{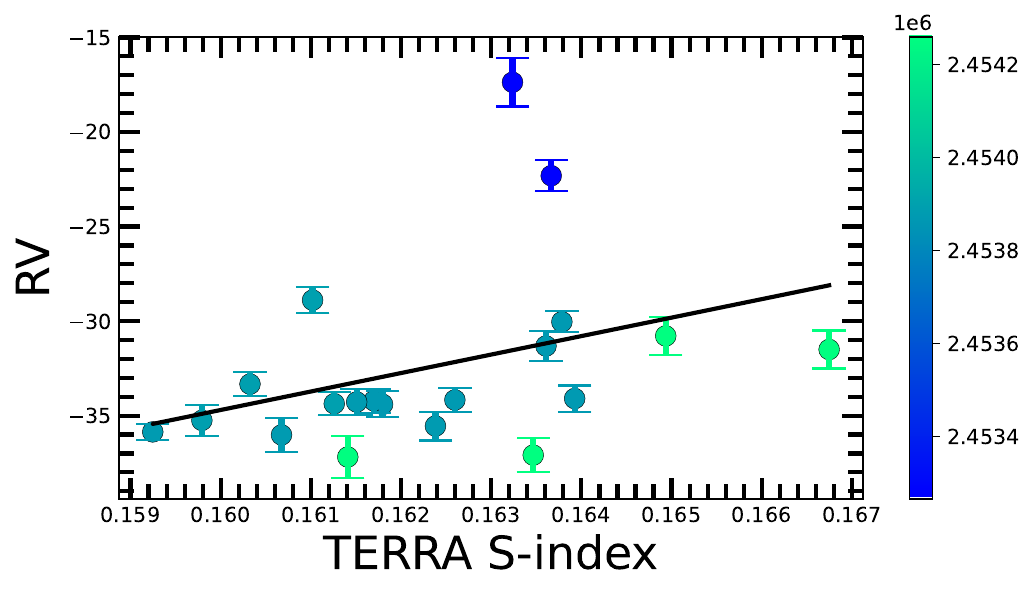}
\includegraphics[width=0.7\columnwidth]{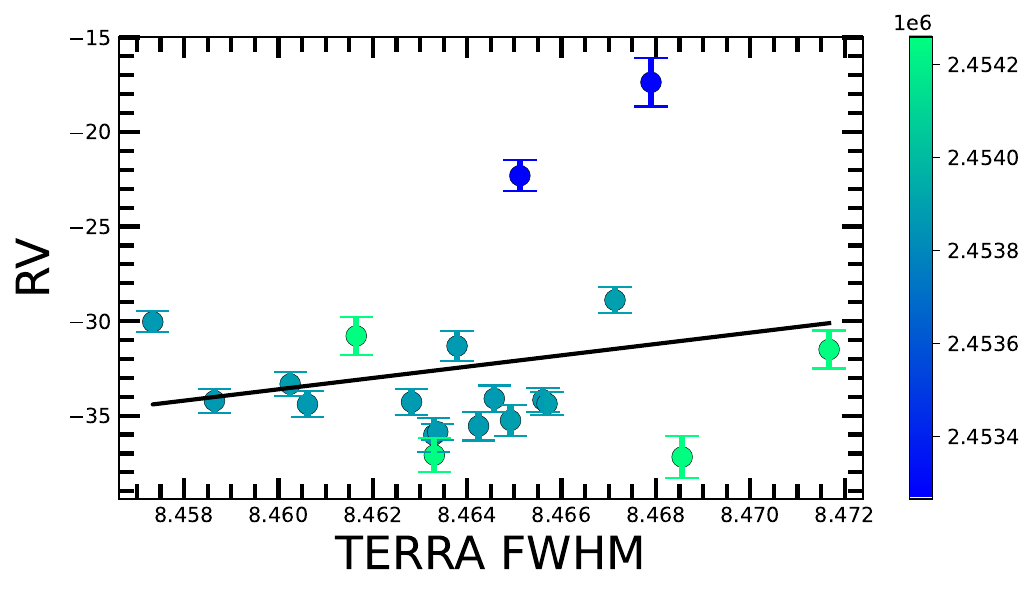}
\includegraphics[width=0.7\columnwidth]{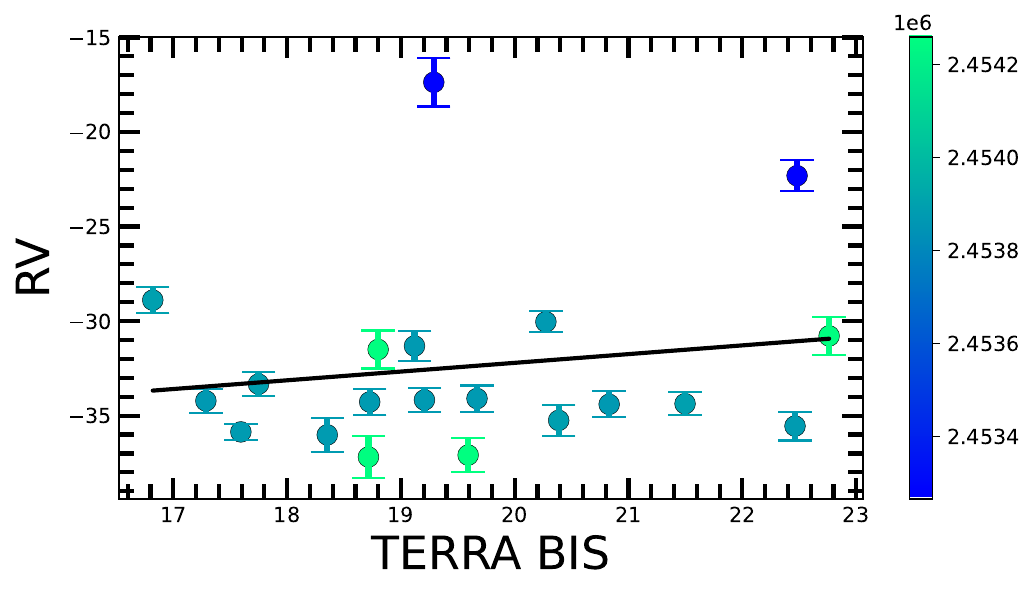}\\
\includegraphics[width=0.7\columnwidth]{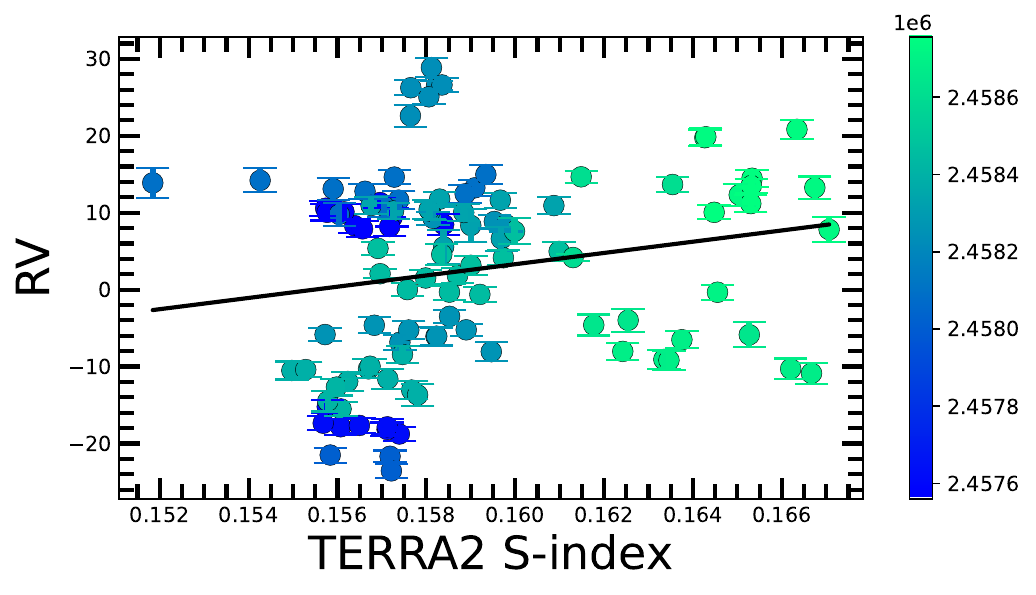}
\includegraphics[width=0.7\columnwidth]{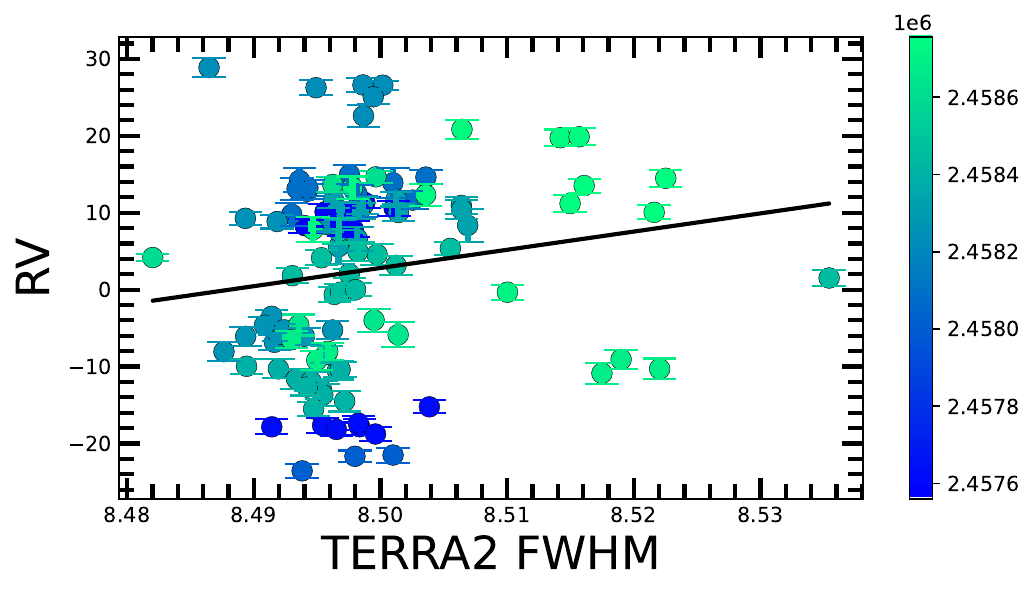}
\includegraphics[width=0.7\columnwidth]{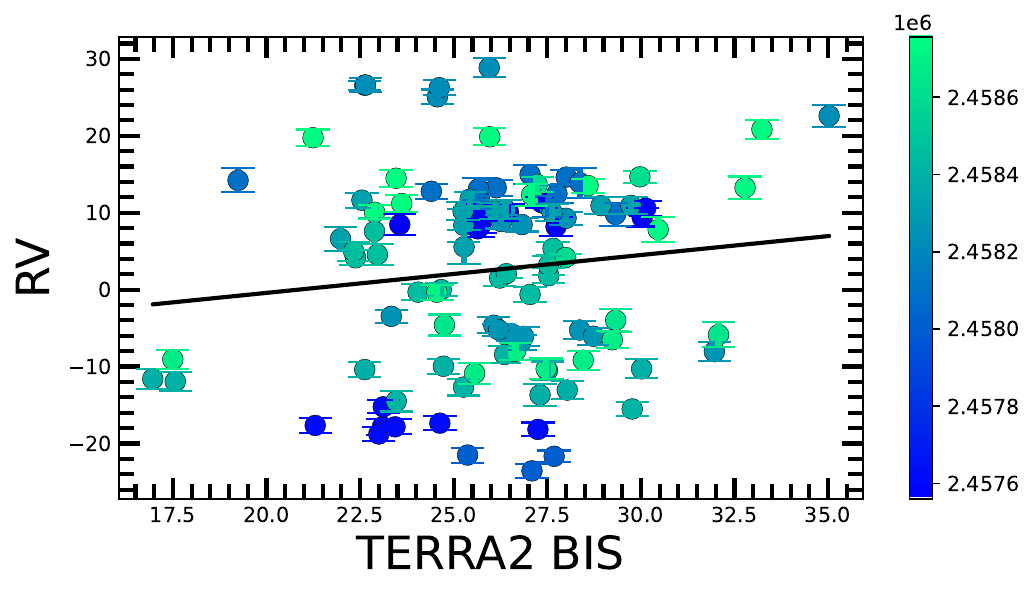}
\caption{Correlations of the available stellar activity indices. From top to bottom, left to right: PFS S-index, TERRA S-index, TERRA FWHM, TERRA BIS, TERRA2 S-index, TERRA2 FWHM, TERRA2 BIS. The lines show the best-fit correlation between the activity indicator and the RVs.}
\label{Activity Correlations}
\end{figure*}

    Following a similar strategy as in \cite{Santos_stellaractivity} and \cite{HD26965Diaz}, we investigated how stellar indices might be affecting or contributing to the periodic signals found within the RVs. First, we checked the GLS periodograms of every available stellar index to see if any stellar activity peak is associated with a previously found periodic signal, or its harmonics, as shown in Figure \ref{Stellar Activity Periodograms}. No signals are significant above the 0.1\% FAP threshold for any index in TERRA or PFS, as well as the Bisector Inverse Slope (BIS) of TERRA2. The S-index in TERRA2 has significant peaks in the mid 20 days and also around 800 days. Also, the Full Width at Half Maximum (FWHM) from TERRA2 has a significant peak at $\sim 22$d. However, none of these peaks approach the previously found periodic signals. The Combined S-index and Combined FWHM both had significant peaks at long periods around $\sim 4000$d. Additionally, the RV vs stellar index correlations were plotted in Figure \ref{Activity Correlations}. Table \ref{Stellar Activity Corr} shows the calculated Pearson rank and p-value of the RV/index correlations. Generally, in order for a correlation to be statistically significant, the p-value must be $p \leq 0.05$. We note that the TERRA S-index, TERRA2 S-index, and TERRA2 FWHM have p-values approaching significance, however none have a rank corresponding to even a moderate correlation. A rank of $\lvert r \rvert \geq 0.5$ is required to be considered a moderate or stronger correlation. Since none of the stellar index correlations have $p \leq 0.05$ nor $ \lvert r \rvert \geq 0.5$, there are no significant nor strong/moderate correlations between the RVs and the stellar indices.

\begin{table}
    \centering
    \caption{Stellar activity index correlations of HD208487}
    \begin{tabular}{lccc}
        \hline \hline Instrument & Index & Pearson rank & p-value \\
        \hline TERRA & S-index & 0.365 & 0.113\\
        TERRA & FWHM & 0.207 & 0.381\\
        TERRA & BIS & 0.163 & 0.492\\
        PFS & S-index & 0.209 & 0.173 \\
        TERRA2 & S-index & 0.189 & 0.052\\
        TERRA2 & FWHM & 0.156 & 0.110\\
        TERRA2 & BIS & 0.122 & 0.212\\
\hline
    \end{tabular}
    
    \label{Stellar Activity Corr}
\end{table}

\subsection{Fitting stellar activity with \emp}
\label{emp_acts_fit}
\begin{table}
    \centering
    \caption{\emp\ results for combined FWHM with $1 \sigma$ errors.}
    \begin{tabular}{lcccc}
        \hline\hline
        & $P$ (days) & $K \ (\mathrm{m/s})$ & $e$ & BIC \\
        \hline
        k1 & $862.3 \genfrac{}{}{0pt}{}{+8.0}{-8.9}$ 
            & $0.011 \genfrac{}{}{0pt}{}{+0.0009}{-0.0008}$ 
            & $0.77 \genfrac{}{}{0pt}{}{+0.02}{-0.01}$ 
            & $-892.99$ \\
        k2 & $43.5\genfrac{}{}{0pt}{}{+0.063}{-0.89}$ 
            & $0.006 \pm 0.0006$ 
            & $0.41 \pm 0.01$ 
            & $-902.66$ \\
        \hline
    \end{tabular}
    \label{emp indices Params}
\end{table}

To expand our search for periodic RV modulations induced by stellar activity, we incorporated Bayesian statistical analysis. As a novel approach, we executed the \emp\ code on the available stellar activity indices (the authors are unaware of any previous procedures involving Bayesian statistical analysis on stellar activity indices).  Utilizing the \emp\ code enables a more rigorous analysis of the stellar activity signals compared with the standard GLS and correlation analysis. Moreover, the posterior distributions derived from the Markov chains afford a comprehensive assessment of certainty across the entire period space, facilitating a thorough evaluation of the distinctness of the signals.\\
\\
We completed several \emp\ runs for each stellar activity index. Each index was initially analyzed individually, before joining the datasets of similar indices from different instruments into one run to test the stability and significance of the signals over the full baseline of the activity data. Further, we tested the differences between how \emp\ fits the stellar activity indices with the fixed zero eccentricity and the constrained ($\mathcal{N} \sim (0,0.3,0,1$)) eccentricity prior. The results of the noteworthy \emp\ run is summarized in Table \ref{emp indices Params}. \\
\\
The result needing further inspection arose from the highest-probability \emp\ run (constrained eccentricity prior) of the combined FWHM (TERRA and TERRA2). The FWHM period found at $P_{FWHM}=860.7\genfrac{}{}{0pt}{}{+8.4}{-9.3}$d somewhat near to the best fit period for HD208487c at $P_{RV}=923.06\genfrac{}{}{0pt}{}{+2.02}{-2.76}$d. Propagating the errors yields a sigma separation of $\sim 7.1\sigma$ away from HD208487c. Thus, based on the current data, the combined FWHM signal is again statistically distinct from the long-period RV signal.  We can therefore say to a high level of statistical confidence that none of the three significant signals we find in the RV data have periods that overlap with signal frequencies in any of the activity indices we analyzed.

\section{Photometric analysis}

    \subsection{ASAS photometry}
\begin{figure*}[t]
   \centering
   \includegraphics[width=\columnwidth]{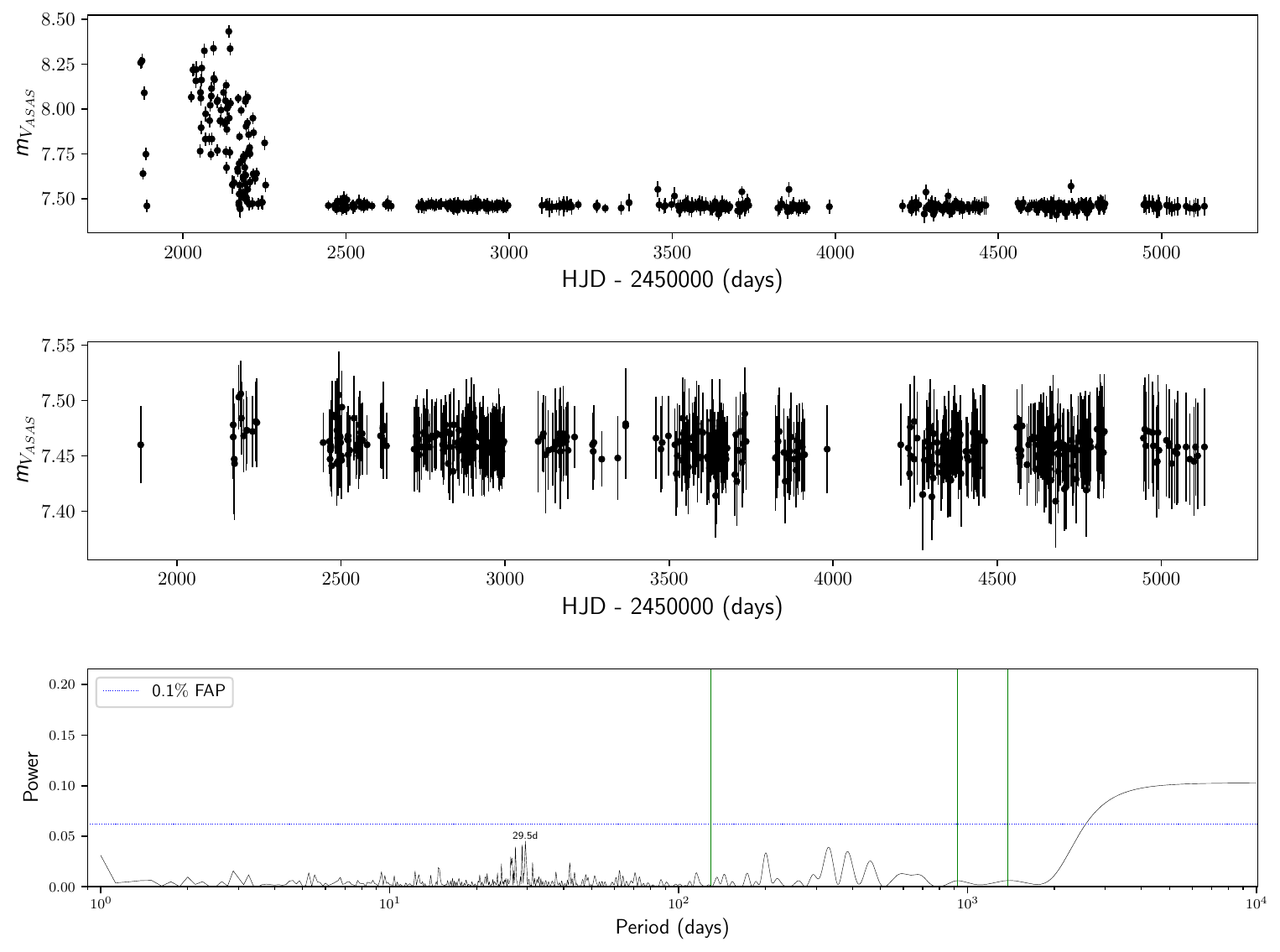}
   \includegraphics[width=\columnwidth]{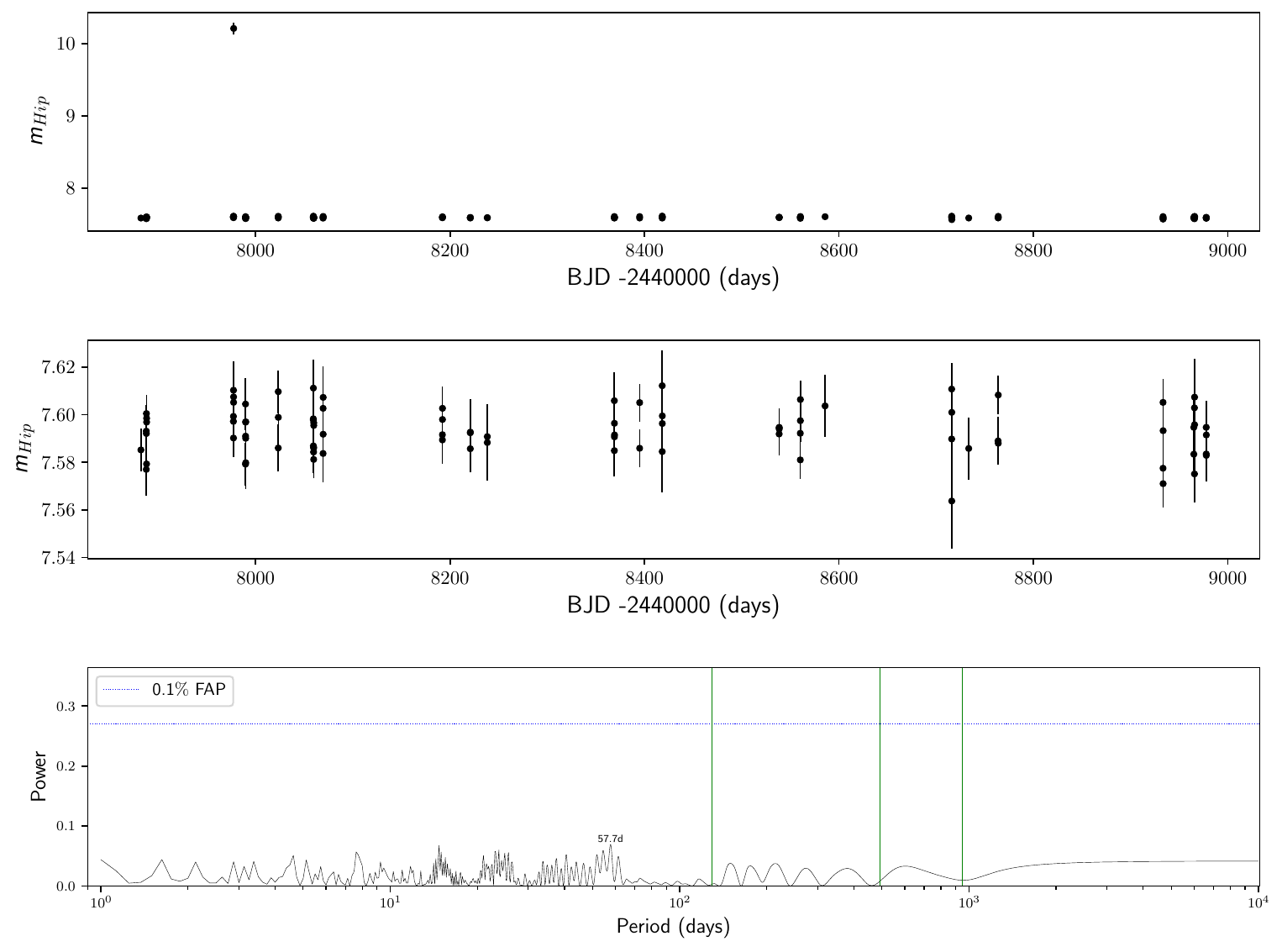}
   \caption{Left (right): ASAS V-band photometry (Hipparcos photometry). Top: "A" and "B" flagged V-band ASAS photometry (all Hipparcos data) for HD208487. Middle: cleaned V-band ASAS (Hipparcos) photometry for HD208487. Bottom: GLS periodogram of the photometry. Horizontal lines correspond to the 0.1\% (dotted blue) FAP significance level computed via 5000 bootstrap iterations on the photometry time series. The vertical green lines represent the periods of the three planets.}
   \label{ASAS Data/Periodogram}
    \end{figure*}

In order to provide additional analysis of this star/planetary system, All-Sky Automated Survey (ASAS) photometric data were analysed \citep{Pojmanski_ASAS}. Searching through the ASAS archive, we found 608 V-band observations spanning just over nine years, between JD 2451870.55 (November 22, 2000) and JD 2455168.58 (December 3, 2009). Following the ASAS guidelines, only data marked with an "A" or "B" quality flagged are kept (henceforth all data), leaving 509 usable data points. However, when examining the photometric time series, we see evidence of instrumental issues in the first $\sim 300$d of observations, see Figure \ref{ASAS Data/Periodogram} top. In order to clean the data, we calculated the mean and standard deviation of all data after JD 2452300. Then all data falling within three standard deviations of the calculated mean for the entire baseline were kept, as seen in Figure \ref{ASAS Data/Periodogram} middle. Finally, the GLS periodogram was calculated from the cleaned data, where we sampled the period space from 1 day to 10,000 days, performing 80,000 period samples. \\
\begin{table}
    \centering
    \caption{The top five peaks in the cleaned ASAS photometry data with their corresponding FAP.}
    \begin{tabular}{lc}
        \hline 
        \hline Period & FAP \\
        \hline
        29.50 & 0.21 \\
        28.75 & 0.39 \\
        27.25 & 0.52 \\
        330.60 & 0.54 \\
        384.47 & 0.84 \\
         \hline
    \end{tabular}

    \label{ASAS Peaks}
\end{table}
\\
Table \ref{ASAS Peaks} shows the top five peaks and their corresponding powers, however all of them fall significantly below the 0.1$\%$ FAP line. It is interesting to note that \cite{HD208487_28d} posit that a $\sim 28-29$d signal they found in earlier RV data could be related to a planet, however given the relatively strong peak around that period in the ASAS data, we believe the planetary hypothesis is unlikely. Although the top two peaks are under the 0.1$\%$ FAP line, they fall in a period range that could be related to either the stellar rotation cycle or the lunar cycle. Additionally, using the same setup, we ran the GLS Periodogram for the entire photometric data set, also showing no significant periods.

\begin{figure}
\includegraphics[width=\columnwidth]{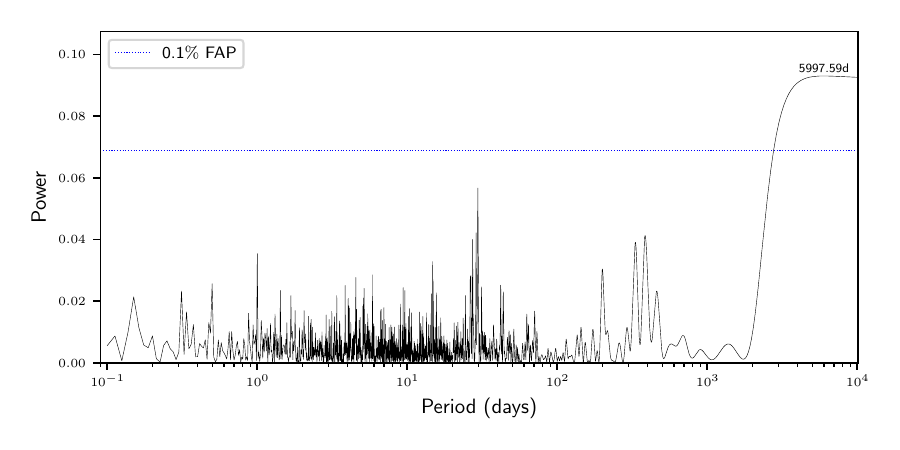} 
\includegraphics[width=\columnwidth]{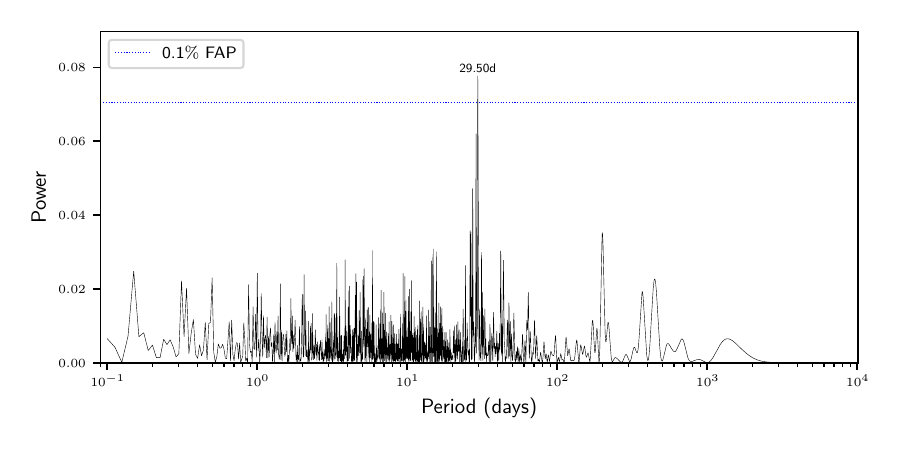}\\
\includegraphics[width=\columnwidth]{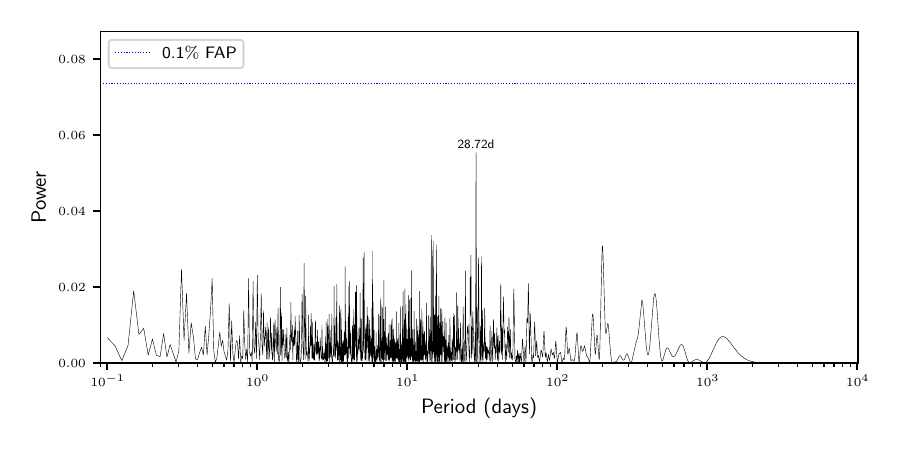}
\caption{GLS periodogram of cleaned ASAS photometry. Top: original GLS periodogram of combined, mean-subtracted ASAS photometry data. Second: GLS periodogram after fully fitting the first signal. Third: GLS periodogram after fully fitting the second signal. The blue, dotted horizontal lines represents the 0.1\% FAP level estimated from 5000 bootsrap resamplings.}
\label{ASAS fits}
\end{figure}
\subsection{Fitting ASAS stellar cycles}

Extending our analysis of the ASAS photometry, we searched for periodic stellar cycles in the cleaned data. After noting that none of the GLS periodogram peaks align with the planetary signals, we fit the strongest periods, as shown in Figure \ref{ASAS fits}. The segment of the periodogram with the highest power is continuously increasing, with no defined peak, even as it approaches a period of 10,000 days. Thus, we conclude that the signal is likely a baseline effect, rather than a real cycle. The next two peaks constitute very similar periods of 29.51 days and 28.73 days. Based on the strong peak in ASAS and a very similar period in the GLS periodogram of the S-index of TERRA2, we conclude that the spikes in this period range are very likely to be due to the stellar rotation period, possibly even being affected by differential rotation.  With continued spectral monitoring of this star we can further test this hypothesis in the future. 

\subsection{Hipparcos photometry}

To further explore the photometric properties of HD208487, we investigated the publicly available Hipparcos data \citep{Hipparcos}. These data were obtained from the NASA Exoplanet Archive\footnote{https://exoplanetarchive.ipac.caltech.edu/}, which contained 90 data points, spanning three years from JD 2447881.99 (December 21, 1989) to JD  2448977.85 (December 21, 1992). Following the data quality flag information, 1 point was identified as low-quality and unusable data. After removing the poor data point, the GLS Periodogram was calculated, in which we sampled the period space from 1 day to 10,000 days, performing 80,000 period samples. There were no peaks coming close to statistical significance, demonstrating a lack of any periodic photometric signals in the Hipparcos data.

\section{Discussion}

    With 184 new RV observations spanning 15+ years since the previous analyses \citep{TinneyDiscoveryHD208487, Gregory_longerP, HD208487_neptune_14/998, Gregoryupdated, HD208487_28d, HD208487_14d}, this system is ripe to study and to re-evaluate the reality of the planetary system. Using the \species\ and \ariadne\ codes, we provide updated stellar parameters to HD208487. In order to fully understand the RV dataset and best characterize the system, we utilized GLS periodograms and \emp\ parallel-tempering MCMC samplings, which revealed new periodic signals embedded in the RVs. An independent, analysis accounting for wavelength-dependent variations in the HARPS-TERRA velocities yielded consistent results providing confidence in the final parameters. The best-fitting model to the current RVs reveals a candidate three-planet system.  \\
\\
Our best-fit model, which confirms the existence of HD208487c, alters some of the parameters. Although the period, $923.06^{+2.02}_{-2.76}$d, falls within the uncertainty published by \citet{Gregory_longerP, Gregoryupdated}, the semi-amplitude, $6.18^{+0.09}_{-0.07}$m/s, and mass, $0.32\pm 0.01 m_j$, are now slightly lower leading to a classification as a Saturn-mass planet. Further, our analysis reveals a third signal, relating to a candidate outer planet, HD208487d, with a period of $1380.13^{+19.20}_{-8.25}$d, a semi-amplitude of $2.46^{+0.30}_{-0.09}$m/s, and a minimum mass of $0.15^{+0.01}_{-0.02}m_j$. The two outer signals are found very close to the 3:2 period ratio ($P_d/P_c = 1.495$), further increasing the reliability of these as being due to Doppler signals, since it is difficult to produce 3:2 frequency ratio signals through aliasing (e.g. \citealp{hara2020}), particularly with a possible 2:1 signal also present in the data, and also transit observations have shown such period ratios to be fairly common (\citealp{lissauer2011,fabrycky2014,lissauer2024}). These tests therefore leave us with a system that presents an interesting and unique system architecture, as there are currently no other known planetary systems that contain two planets on 500-1500d orbits that are between $0.1-0.4m_j$. HD208487 contains an inner gas giant along with a resonant pair of outer, Saturn/sub-Saturn planets that are close to a second order period resonance. \\
\\
As part of our analysis, we examined all of the available stellar activity indices, including, to our knowledge, the first use of Bayesian analysis on stellar activity indices to check for activity-induced RV signals. We find that none of the stellar indices have peaks matching the RV periods of the discovered planets, nor their harmonics. Although the Bayesian analysis of the stellar activity indices revealed one signal (Combined FWHM) in proximity to the RV signals, it is statistically distinct based on the current data. Additionally, none of the stellar activity indices are statistically significantly correlated with the observed RVs, nor reach the level of even a moderate correlation, which was not the case for analyses done on stars like HD26965, for example \citet{HD26965Diaz}. Furthermore, during the GLS periodogram analysis with \exo, we split the RV data sets into two separate sets, based on their S-index values. Although only the inner gas giant was detected in the active dataset, three signals were found in the inactive dataset, the inner gas giant, the $\sim 500d$ signal, and the Saturn-mass planet, HD208487c. Since the model fit of the inactive set was significantly better, the stellar activity may actually be masking the Doppler signals rather than contributing to them. Again, we reiterate that more data is needed to confirm the nature of the intermediate, $\sim 500d$ signal.\\
\\
In order to evaluate the stability of the signals over time, we split the data into early and later subsets. HD208487b remains entirely consistent in both data sets. HD208487c is also found in both data sets, with the period and semi-amplitude remaining consistent within uncertainties. The major difference between the earlier and later data sets relates to the period of the third significant signal. The earlier data yields the longest-period signal, while an intermediate-period signal arises in the later data. Both the overall best-fit \emp\ model and an independent analysis favor the longest-period signal, leading to the adoption of HD208487d as an outer planet.  However, we note that often times our Bayesian analyses arrived at a model favoring the solution with the shorter-period planet, yet even in the best case scenarios the model probabilities were far below the models containing the outer signal.  Yet this might indicate that there is indeed another Doppler signal close to a 2:1 period ratio with the HD208487c signal, and from a dynamical standpoint this also may be favored, since it would be another example of a dynamically packed planetary system.\\
\\
Extending our investigation of this system, we checked ASAS photometry data to search for any periodic variations in the star's visible magnitude. Examining the data showed no significant peaks within the photometry. However, the highest power peaks, paired with stellar activity indices from TERRA2, suggest a stellar rotational period around $P_{rot} \sim 29$d. 
\subsection{Previous signal conclusions}
After the discovery of HD208487b by \cite{TinneyDiscoveryHD208487} and the subsequent detection of HD208487c by \cite{Gregory_longerP}, some disagreements about the planetary configuration arose. \cite{Gregoryupdated}, \cite{HD208487_neptune_14/998}, and \cite{HD208487_28d} also detected HD208487c as a moderately eccentric ($e \simeq 0.4$), sub-Jovian world, however the latter two papers find equally probable solutions at $P_c \sim 14.5$d, $P_c \sim 28$d, or as an activity-induced RV signal, leading them to disregard the existence of HD208487c as an outer planet. \cite{HD208487_14d} also claim a potential planet around $P_c \sim 14.5$d. \\
\\
Conclusions drawn from our analysis, applying updated analysis methods and much more data over a longer baseline and at high RV precision, posits that the stellar rotation period is around $P_{rot} \sim 29$d, possibly with some evidence for differential rotation, and therefore this leads to the conclusion that \cite{HD208487_neptune_14/998, HD208487_28d, HD208487_14d} actually found the rotation period and a likely alias at half this rotation period. Further, we believe that the detection of HD208487c as a moderately eccentric, sub-Jovian world fits the original data, as seen in Section \ref{AAT-only}. The new, higher precision RV data collected over the last 15+ years now enables the detection of candidates HD208487d, which would be an outer super-Neptune/sub-Saturn planet, and HD208487c, which would be a Saturn-mass planet in a near-circular orbit, explaining the different planetary models discussed in these works.
\subsection{Habitable zone planets}
As we have found new planets in this work, we can try to more deeply understand and characterize them. Without observing a transit, we do not have a way to constrain their radii, meaning we can't determine their densities or physical parameters. However, since their orbits are well-constrained, we can gain a better understanding of this multi-planet system. Based on the definition described by \citet{HZMainSequence} and \citet{HZ_new}, we can measure the limits of the Habitable Zone (HZ) of HD208487, using the following equations:
\begin{equation}
        S_{eff} = S_{eff \odot} + aT_{\star} + bT^2_{\star} + cT^3_{\star} + dT^4_{\star}        
\end{equation}
where $S_{eff}$ is the effective solar flux, $T_{\star} = T_{eff} - 5780K$ ($T_{eff}$ is the stellar effective temperature), and a, b, c, d, and $S_{eff \odot}$ are constants whose values are given in Table 3 of \cite{HZ_new}.
\begin{equation}
    d = \sqrt{\frac{L_{\star} \slash L_{\odot}}{S_{eff}}}
\end{equation}
where $L_{\star}$ and $L_{\odot}$ is the luminosity of HD208487, in this case, and the Sun, respectively, and the radius, d, is measured in astronomical units (AU). We used the coefficient values from the "recent Venus" and "early Mars" models \citep{HZ_new}. From the results of the \ariadne\ code, the luminosity of HD208487 is $L_\star = 1.71L_\odot$. As a result we found that the HZ ranges from 0.97AU to 2.25AU. We find that HD208487c would orbit the star at around 1.94AU and HD208487d would orbit at around 2.54AU, so HD208487c would lie within the HZ, while HD208487d would orbit just outside the proposed range of the HZ  \citep{Habitability208487}. 

\subsection{Exomoons}
Long period exoplanets are also interesting as they could potentially host natural satellites (also known as exomoons). This is because, for the case of planets in close-in orbits, the extreme gravitational interactions with the host stars could destabilize potential exomoon orbits \cite[e.g.][]{2010ApJ...719L.145N, 2016ApJ...817...18S, 2021PASP..133i4401D}. Moons are thought to be formed through one of two processes, i.e. coaccretion in the circumplanetary discs around the young planets \cite[e.g.][]{2014AsBio..14..798H, 2016AstRv..12...24B, 2022NatCo..13..568N} and planetary collisions/captures \citep{2014AsBio..14..798H, 2016AstRv..12...24B, 2017MNRAS.466.4868B}. Since coaccretion in the circumplanetary discs might require larger amount of materials to form moons, this process might be possible in the case of larger Jupiter-sized planets, as could be the case with HD208487c. However, for HD208487d, which appears to be a Neptunian-sized, planetary collisions/capture could potentially lead to a moon formation.\\
\\
The critical distance from the host planet, within which a moon could have a stable orbit, depends upon the Hill radius of the planet, the eccentricity of the planet’s orbit, and the relative inclination of the moon’s orbit (i.e. if the moon’s orbit is prograde or retrograde with respect to the planet). This critical orbital distance can be expressed as $0.4895$R$_H(1-1.0305e)$ for a prograde orbit of the moon, and $0.9309$R$_H(1.0000-1.0764e)$  for a retrograde orbit of the moon \citep{2006MNRAS.373.1227D, 2010ApJ...719L.145N}, where R$_H$ is the Hill radius, and $e$ is the orbital eccentricity of the host planet.\\
\\
Since both HD208487c and HD208487d signals have been detected only through the RV method, we can only estimate the minimum mass, and thus, the minimum radius of the Hill sphere for them, which we have estimated to be 12.935 MKm and 13.156 MKm respectively. The minimum distance from the host-star within which a moon can survive in a prograde orbit around HD208487c and HD208487d are 4.993 MKm and 5.599 MKm respectively, and those for a retrograde orbit are 9.57 MKm and 10.786 MKm respectively. For comparison, the distance of the outermost Galilean moon, Callisto, from Jupiter is 1.883 MKm, and the largest retrograde satellite in our Solar system, Triton, orbits around Neptune only at a distance of 0.354 MKm. Thus, both HD208487c and HD208487d are capable of hosting potential exomoons, and since they are both orbiting in the HZ around their host star, potential presence of HZ exomoons make them even more interesting. However, detecting such exomoons would require extremely high precision photometric and spectroscopic follow-ups, which could only be possible using the next generation large telescopes \citep[e.g.][]{2009EM&P..105..385S, 2009MNRAS.392..181K, 2020A&A...635A..59T, 2022ApJ...936....2S, saha2022precise, 2024BSRSL..93..123S}.
\subsection{Dynamics}

\begin{figure}
    \centering    
    \includegraphics[width=\columnwidth]{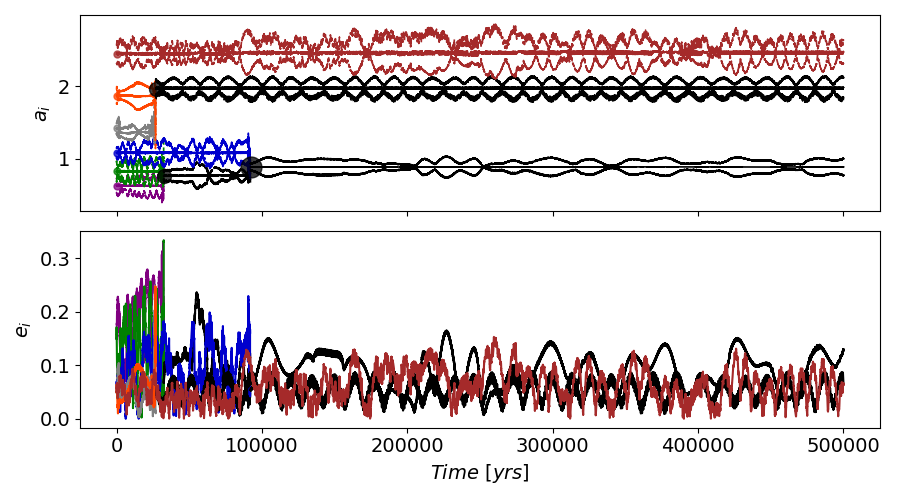}
    \caption{Temporal evolution of a six-planet simulation showing a merger between the first three and the fourth and fifth planets, leaving behind a 3 planet system with similar masses to the observations. The upper panel displays the semi-major axis $a_i$, the variation of the apocenter $Q$ (defined as $a_i[1+e_i]$), and the pericenter ($q=a_i[1-e_i]$). In the bottom panel we show the variation of the eccentricities. }
    \label{fig:collision}
\end{figure}

Motivated by the apparent architecture of the HD208487 system, and with the aim of testing if such a system could be possible and understandable, we explore the possibility that the system's origin could be attributed to a dynamical instability of an initially more ordered system. The star hosts two moderately eccentric inner planets ($e_1\sim0.37$ and $e_2\sim0.2$), each three and two times more massive than the outer Neptune-like planet. The two outer planets are near resonance, while the innermost giant appears dynamically detached from the outer two.  \\
\\
We posit that this more ordered system consisted of a coplanar six-planet system orbiting a $M_\star = 1.15 M_\odot$ central star in a near-resonant chain configuration (with consecutive period ratios near 3:2). We envision a scenario in which a six-planet system was formed in a gaseous disk and got captured in resonance due to planet-disk interactions. After dispersal of the disk, the resonance chain is broken and the system becomes unstable---a scenario reminiscent to breaking of the resonance chains applied to super-Earths in the Kepler sample \citep{2017MNRAS.470.1750I}.  However, unlike the Kepler planets where instabilities mainly lead to mergers between planets, the cold super-Neptunes in HD208487 may lead to either collisions between planets or ejections from the system. We can quantify their relative probabilities using the so-called Safronov number for parabolic orbits as
\begin{eqnarray}
    \theta^2 &\equiv& \left(\frac{2GM_p}{R_p}\right)\left(\frac{a_p}{GM_\star}\right)\nonumber\\
    &=&\left( \frac{M_p}{M_{\rm J}} \frac{R_{\rm Jup}}{R_p} \frac{a_p}{0.25 {\rm au}} \right)
              \left( \frac{M_\odot}{M_\star} \right),
    \label{eq:safronov}
\end{eqnarray}
where for $\theta^2 \gg 1$ the planetary radius is so small that close encounters mostly lead to ejections relative to collisions, while for $\theta^2<1 $ collisions are more frequent \citep{2001Icar..150..303F,2014ApJ...786..101P}. 
A typical super-Earth at 0.5 au has $\theta^2\sim 0.25$ leading to collisions and small eccentricity excitation \citep{2001Icar..150..303F,2014ApJ...786..101P}, but a cold Neptune at $\sim 2$ au has  $\theta^2\sim 1$ leading to mixed outcomes and potential mergers with significant eccentricity excitation. We numerically evaluate this possibility  next.\\
\\
We perform 250 simulations using the REBOUND code \citep{2012A&A...537A.128R} with the primary objective of investigating the possibility of achieving an orbital configuration resembling the observed data, thereby deviating from the conventional `peas in a pod' scenario. The planetary masses are $m_i = 0.15 \, m_{\rm Jup}$ with $i=1\dots 6$, and we place the planets 5\% away from the exact 3:2 - 3:2 - 3:2 - 3:2 - 3:2 resonant chain.
We chose 5 values for the initial eccentricities for the two inner planets: $e_1 = e_2 = 0.05, 0.1, 0.15, 0.2, 0.25$ to speed up the phase of dynamical instability, while the outer four bodies start with a fixed value of 0.05. For each of them we randomly selected 50 values for the initial argument of periastron $\omega_i \in [0, 2.\pi]$ and inclinations randomly distributed between $0^\circ$ and $2^\circ$. Collisions happen whenever the distance between two planets is less than the sum of their radii: $d < r_i + r_j$, where the radius of each planet follows the $r_i \propto m_i^{-2.06}$ law by \citet{lissauer2011}.\\
\\
Over a simulation period of $5 \times 10^5$ years, none
of the systems remained stable and maintained their six-planet configuration. In this compact configuration, all the simulations underwent an instability leading to planet collisions generally leaving behind 3, 4, and 5 planets. We note that the rate of stable (unstable) systems is highly sensitive to the initial conditions and will likely decrease (increase) as our simulations continue to evolve for longer timescales. Around $50\%$ of the systems that underwent an instability, led to the formation of a three-planet system. Of these, in $\sim20\%$ of cases, the three inner planets merged as a massive innermost body, and the fourth and fifth planets merged as an intermediate mass middle planet, resulting in a three-planet system with an architecture similar to the one observed in our data, with inclinations reaching a final root mean square (RMS) of $1^\circ$. Figure \ref{fig:collision} illustrates the temporal evolution of a representative simulation where, at $\sim5 \times 10^4$ years, a collision occurs leading to the merging of the inner 2 bodies into a single planet, after which a  second collision occurs at $\sim 10^5$ years, finally forming the innermost body, $m_1$. Around $\sim5 \times 10^4$ years, a collision of the $4^{th}$ and $5^{th}$ forms the middle planet, $m_2$, while the outermost body never collides with other bodies. The top panel displays the semi-major axis, peri- and apocentric distances of each planet ($a_i(1\mp e_i)$, respectively), while the bottom panel showcases the eccentricities as a function of time. The post-merger planet attains high eccentricities, providing a potential explanation for the observed eccentricities of the inner planets.\\
\\
Although the approximation of treating collisions as perfectly inelastic mergers might not capture the complete realism of the scenario, it provides valuable insights as a proof of concept for understanding how such a system can form. As already stated by \citet{2017MNRAS.470.1750I}, once the gas dissipates resonant chains may become dynamically unstable, undergoing a phase of gravitational scattering. Our simulations begin after the dissipation of the gas, exploiting the phase of dynamical instability, allowing the formation of a system featuring a fairly eccentric massive inner planet, largely separated from the outer pair.

\section{Conclusions}
    
    This paper provides a thorough re-investigation into the HD208487 planetary system. First, we derived an updated set of stellar parameters utilizing the stellar modeling codes, \species\ and \ariadne. Next, we comprehensively analyzed the currently available RV data of HD208487 to search for signals not previously found in older investigations. Further, we developed a more rigorous methodology to analyze stellar activity indices. Finally, we evaluated ASAS and Hipparcos data to search for any periodic photometric signals. \\
\\
As a result of our RV analysis, we determined that HD208487 possibly contains at least three planets in the system (see Table \ref{Final Params}). We provided updated parameters to the sub-Jovian HD208487b with a period of $129.36$d.  Formerly believed to be a possible sub-Jovian and moderately eccentric planet (\citealp{Gregory_longerP}), we now show HD208487c is actually a candidate Saturn with a period of $923.06$d and a near-circular orbit.  Additionally, we discovered a new super-Neptune/sub-Saturn candidate, HD208487d, a planet that would be orbiting with a period of $1380.13$d. HD208487c and HD208487d have an orbital period ratio of $P_d/P_c = 1.495$, indicating they are very close to a 3:2 MMR.  There is also some evidence that leads us to believe that there may exist a 4th planet in this system at close to the 2:1 period ratio with HD208487c, which would further dynamically pack the planetary system, however more long-baseline RV data will be required to confirm this as a reality or not.\\
\\
Our methodology to analyze the available stellar activity indices with a Bayesian statistical analysis enabled us to distinguish the RV signals from the statistically significant activity signals. We found no correlations between the RVs and stellar activity signals and no peaks in the GLS at the periods of the three planets. Using \emp\ to provide a more accurate description, we found that the nearest detected stellar activity signal is separated by $7.1\sigma$ from $P_c$. Thus, we determined that there is a high probability that the RVs and stellar activity signals are distinct, based on the current data.  An independent analysis of the RVs using colour-dependent terms to detrend sources of colour dependent noise also did not yield any correlations, arriving at the same system configuration as that of the EMPEROR analysis.\\
\\
Concluding our observational analysis, we studied the ASAS and Hipparcos photometric data. After fitting the long-period, baseline-effect signal, we found a statistically significant period at $\sim 29$d. Combining this result with our GLS and \emp\ analysis of the stellar activity indices, we believe this signal to indicate the rotation period of the star, with potential differential rotation. This result explains the false RV signals, and a likely half-period alias, posited by \cite{HD208487_neptune_14/998}, \cite{HD208487_28d}, and \cite{HD208487_14d}. \\
\\
Finally, we performed dynamical simulations in an effort to better understand the formation history of this complex system.  By assuming the common  `peas in a pod' configuration to begin with, including additional planets in the system close to a resonant chain, we were able to show that the inner moderately eccentric gas giant could be the remnant of a previous collision of lower-mass planets.  
Of the systems that underwent an instability and subsequent planetary collisions, such a scenario occurred $\sim20\%$ of the time in our simulation set.   
This scenario originates from systems with Safranov numbers close to unity, that allows for significant scattering and eccentricity excitation before the merger occurs, possibly representing the breaking point of more ordered configurations and providing a chaotic look at the final architectures of planetary systems. Systems like HD208487 in our simulations, end up with relatively low inclinations ($RMS\sim 1^\circ$), similar to the inclinations of Kepler planets within 3-planet systems \citep{fabrycky2014,2018ApJ...860..101Z}. However, unlike Kepler planets, HD208487 planets orbit with semi-major axes $>1 $ au, unreachable by Kepler. Future missions like PLATO may spot systems with these architectures. 
\section*{Data Availability}
Tables~A.1-A.4 are only available in electronic form at the CDS via anonymous ftp to 
cdsarc.u-strasbg.fr (130.79.128.5) or via 
\url{http://cdsweb.u-strasbg.fr/cgi-bin/qcat?J/A+A/}.

\begin{acknowledgements}
      Part of this work was supported by the United States Fulbright Fellowship in partnership with the Fulbright Chile Commission. JSJ gratefully acknowledges support by FONDECYT grant 1201371, from the ANID BASAL project FB210003, and from CASSACA grant CCJRF2205. CC was supported by FONDECYT grant n$^\circ$ 3230283. CP acknowledges support from CATA-Basal AFB-170002, ANID BASAL project FB210003, FONDECYT Regular grant 1210425, CASSACA grant CCJRF2105, and ANID+REC Convocatoria Nacional subvencion a la instalacion en la Academia convocatoria 2020 PAI77200076. MT acknowledges support from the Jenny and Antti Wihuri Foundation. This work has also made use of the University of Hertfordshire's high-performance computing facility.
\end{acknowledgements}

\bibliographystyle{aa} 

\bibliography{HD208487}

\begin{thebibliography}{115}
\expandafter\ifx\csname natexlab\endcsname\relax\def\natexlab#1{#1}\fi

\bibitem[{Hip(1997)}]{Hipparcos}
 1997, ESA Special Publication, Vol. 1200, {The HIPPARCOS and TYCHO catalogues. Astrometric and photometric star catalogues derived from the ESA HIPPARCOS Space Astrometry Mission}

\bibitem[{{Allard} {et~al.}(2012){Allard}, {Homeier}, \& {Freytag}}]{Allard2012}
{Allard}, F., {Homeier}, D., \& {Freytag}, B. 2012, Philosophical Transactions of the Royal Society of London Series A, 370, 2765

\bibitem[{Anglada-Escud{\'e} {et~al.}(2016)Anglada-Escud{\'e}, Amado, Barnes, Berdi{\~n}as, Butler, Coleman, de~la Cueva, Dreizler, Endl, Giesers, Jeffers, Jenkins, Jones, Kiraga, K{\"u}rster, L{\'o}pez-Gonz{\'a}lez, Marvin, Morales, Morin, Nelson, Ortiz, Ofir, Paardekooper, Reiners, Rodr{\'\i}guez, Rodrίguez-L{\'o}pez, Sarmiento, Strachan, Tsapras, Tuomi, \& Zechmeister}]{ProximaCentaurib}
Anglada-Escud{\'e}, G., Amado, P.~J., Barnes, J., {et~al.} 2016, Nature, 536, 437

\bibitem[{{Anglada-Escud{\'e}} {et~al.}(2012){Anglada-Escud{\'e}}, {Arriagada}, {Vogt}, {Rivera}, {Butler}, {Crane}, {Shectman}, {Thompson}, {Minniti}, {Haghighipour}, {Carter}, {Tinney}, {Wittenmyer}, {Bailey}, {O'Toole}, {Jones}, \& {Jenkins}}]{GJ667}
{Anglada-Escud{\'e}}, G., {Arriagada}, P., {Vogt}, S.~S., {et~al.} 2012, \apjl, 751, L16

\bibitem[{{Anglada-Escude} \& {Butler}(2012)}]{TERRA}
{Anglada-Escude}, G. \& {Butler}, R.~P. 2012, VizieR Online Data Catalog, J/ApJS/200/15

\bibitem[{{Anglada-Escud{\'e}} {et~al.}(2013){Anglada-Escud{\'e}}, {Tuomi}, {Gerlach}, {Barnes}, {Heller}, {Jenkins}, {Wende}, {Vogt}, {Butler}, {Reiners}, \& {Jones}}]{GJ667_Revisited}
{Anglada-Escud{\'e}}, G., {Tuomi}, M., {Gerlach}, E., {et~al.} 2013, \aap, 556, A126

\bibitem[{{Babu} {et~al.}(2010){Babu}, {Stoica}, {Li}, {Chen}, \& {Ge}}]{HD208487_14d}
{Babu}, P., {Stoica}, P., {Li}, J., {Chen}, Z., \& {Ge}, J. 2010, \aj, 139, 783

\bibitem[{Barnes {et~al.}(2020)Barnes, Haswell, Staab, Anglada-Escud{\'e}, Fossati, Doherty, Cooper, Jenkins, D{\'\i}az, Soto, \& Pe{\~n}a~Rojas}]{DMPP3_emp4}
Barnes, J.~R., Haswell, C.~A., Staab, D., {et~al.} 2020, Nature Astronomy, 4, 419

\bibitem[{{Barr}(2016)}]{2016AstRv..12...24B}
{Barr}, A.~C. 2016, The Astronomical Review, 12, 24

\bibitem[{{Barr} \& {Bruck Syal}(2017)}]{2017MNRAS.466.4868B}
{Barr}, A.~C. \& {Bruck Syal}, M. 2017, \mnras, 466, 4868

\bibitem[{Beaug{\'e} {et~al.}(2006)Beaug{\'e}, Michtchenko, \& Ferraz-Mello}]{Migration2/1}
Beaug{\'e}, C., Michtchenko, T.~A., \& Ferraz-Mello, S. 2006, Monthly Notices of the Royal Astronomical Society, 365, 1160

\bibitem[{Butler \& Marcy(1996)}]{47UrsaeMajoris}
Butler, R.~P. \& Marcy, G.~W. 1996, The Astrophysical Journal, 464, L153

\bibitem[{{Butler} {et~al.}(1996){Butler}, {Marcy}, {Williams}, {McCarthy}, {Dosanjh}, \& {Vogt}}]{IodineMethod}
{Butler}, R.~P., {Marcy}, G.~W., {Williams}, E., {et~al.} 1996, \pasp, 108, 500

\bibitem[{{Butler} {et~al.}(2001){Butler}, {Tinney}, {Marcy}, {Jones}, {Penny}, \& {Apps}}]{AAPS}
{Butler}, R.~P., {Tinney}, C.~G., {Marcy}, G.~W., {et~al.} 2001, \apj, 555, 410

\bibitem[{{Butler} {et~al.}(2017){Butler}, {Vogt}, {Laughlin}, {Burt}, {Rivera}, {Tuomi}, {Teske}, {Arriagada}, {Diaz}, {Holden}, \& {Keiser}}]{butler17}
{Butler}, R.~P., {Vogt}, S.~S., {Laughlin}, G., {et~al.} 2017, \aj, 153, 208

\bibitem[{{Butler} {et~al.}(2006){Butler}, {Wright}, {Marcy}, {Fischer}, {Vogt}, {Tinney}, {Jones}, {Carter}, {Johnson}, {McCarthy}, \& {Penny}}]{Butler208487}
{Butler}, R.~P., {Wright}, J.~T., {Marcy}, G.~W., {et~al.} 2006, \apj, 646, 505

\bibitem[{{Castelli} \& {Kurucz}(2003)}]{ATLAS9}
{Castelli}, F. \& {Kurucz}, R.~L. 2003, in Modelling of Stellar Atmospheres, ed. N.~{Piskunov}, W.~W. {Weiss}, \& D.~F. {Gray}, Vol. 210, A20

\bibitem[{{Chatterjee} {et~al.}(2008){Chatterjee}, {Ford}, {Matsumura}, \& {Rasio}}]{2008ApJ...686..580C}
{Chatterjee}, S., {Ford}, E.~B., {Matsumura}, S., \& {Rasio}, F.~A. 2008, \apj, 686, 580

\bibitem[{Costes {et~al.}(2021)Costes, Watson, de~Mooij, Saar, Dumusque, Cameron, Phillips, G{\"u}nther, Jenkins, Mortier, \& Thompson}]{LongtermStellarActivity_RVvariations}
Costes, J.~C., Watson, C.~A., de~Mooij, E., {et~al.} 2021, Monthly Notices of the Royal Astronomical Society, 505, 830

\bibitem[{Crane {et~al.}(2010)Crane, Shectman, Butler, Thompson, Birk, Jones, \& Burley}]{PFS_new}
Crane, J., Shectman, S., Butler, R., {et~al.} 2010, Proceedings of SPIE - The International Society for Optical Engineering

\bibitem[{{Crane} {et~al.}(2006){Crane}, {Shectman}, \& {Butler}}]{PFS}
{Crane}, J.~D., {Shectman}, S.~A., \& {Butler}, R.~P. 2006, in Society of Photo-Optical Instrumentation Engineers (SPIE) Conference Series, Vol. 6269, Society of Photo-Optical Instrumentation Engineers (SPIE) Conference Series, ed. I.~S. {McLean} \& M.~{Iye}, 626931

\bibitem[{{Crane} {et~al.}(2008){Crane}, {Shectman}, {Butler}, {Thompson}, \& {Burley}}]{Crane2008}
{Crane}, J.~D., {Shectman}, S.~A., {Butler}, R.~P., {Thompson}, I.~B., \& {Burley}, G.~S. 2008, in Society of Photo-Optical Instrumentation Engineers (SPIE) Conference Series, Vol. 7014, Ground-based and Airborne Instrumentation for Astronomy II, ed. I.~S. {McLean} \& M.~M. {Casali}, 701479

\bibitem[{Dawson {et~al.}(2019)Dawson, Huang, Lissauer, Collins, Sha, Armstrong, Conti, Collins, Evans, Gan, Horne, Ireland, Murgas, Myers, Relles, Sefako, Shporer, Stockdale, {\v Z}erjal, Zhou, Ricker, Vanderspek, Latham, Seager, Winn, Jenkins, Bouma, Caldwell, Daylan, Doty, Dynes, Esquerdo, Rose, Smith, \& Yu}]{TOI216_WarmResonance}
Dawson, R.~I., Huang, C.~X., Lissauer, J.~J., {et~al.} 2019, The Astronomical Journal, 158, 65

\bibitem[{D{\'\i}az {et~al.}(2020)D{\'\i}az, Jenkins, Feng, Butler, Tuomi, Shectman, Thorngren, Soto, Vines, Teske, Dragomir, Villanueva, Kane, Berdi{\~n}as, Crane, Wang, \& Arriagada}]{MagellanPFS_Diaz}
D{\'\i}az, M.~R., Jenkins, J.~S., Feng, F., {et~al.} 2020, Monthly Notices of the Royal Astronomical Society, 496, 4330

\bibitem[{{D{\'\i}az} {et~al.}(2018){D{\'\i}az}, {Jenkins}, {Tuomi}, {Butler}, {Soto}, {Teske}, {Feng}, {Shectman}, {Arriagada}, {Crane}, {Thompson}, \& {Vogt}}]{HD26965Diaz}
{D{\'\i}az}, M.~R., {Jenkins}, J.~S., {Tuomi}, M., {et~al.} 2018, \aj, 155, 126

\bibitem[{Diego {et~al.}(1990)Diego, Charalambous, Fish, \& Walker}]{UCLES}
Diego, F., Charalambous, A., Fish, A.~C., \& Walker, D.~D. 1990, in Instrumentation in Astronomy VII, ed. D.~L. Crawford, Vol. 1235, International Society for Optics and Photonics (SPIE), 562 -- 576

\bibitem[{{Dobos} {et~al.}(2021){Dobos}, {Charnoz}, {P{\'a}l}, {Roque-Bernard}, \& {Szab{\'o}}}]{2021PASP..133i4401D}
{Dobos}, V., {Charnoz}, S., {P{\'a}l}, A., {Roque-Bernard}, A., \& {Szab{\'o}}, G.~M. 2021, \pasp, 133, 094401

\bibitem[{{Domingos} {et~al.}(2006){Domingos}, {Winter}, \& {Yokoyama}}]{2006MNRAS.373.1227D}
{Domingos}, R.~C., {Winter}, O.~C., \& {Yokoyama}, T. 2006, \mnras, 373, 1227

\bibitem[{{Dotter}(2016)}]{MESA/MIST}
{Dotter}, A. 2016, \apjs, 222, 8

\bibitem[{{Dumusque} {et~al.}(2011){Dumusque}, {Lovis}, {S{\'e}gransan}, {Mayor}, {Udry}, {Benz}, {Bouchy}, {Lo Curto}, {Mordasini}, {Pepe}, {Queloz}, {Santos}, \& {Naef}}]{MagCycles_ActivityVariation}
{Dumusque}, X., {Lovis}, C., {S{\'e}gransan}, D., {et~al.} 2011, \aap, 535, A55

\bibitem[{{Fabrycky} {et~al.}(2014){Fabrycky}, {Lissauer}, {Ragozzine}, {Rowe}, {Steffen}, {Agol}, {Barclay}, {Batalha}, {Borucki}, {Ciardi}, {Ford}, {Gautier}, {Geary}, {Holman}, {Jenkins}, {Li}, {Morehead}, {Morris}, {Shporer}, {Smith}, {Still}, \& {Van Cleve}}]{fabrycky2014}
{Fabrycky}, D.~C., {Lissauer}, J.~J., {Ragozzine}, D., {et~al.} 2014, \apj, 790, 146

\bibitem[{{Feng} {et~al.}(2017){Feng}, {Tuomi}, {Jones}, {Barnes}, {Anglada-Escud{\'e}}, {Vogt}, \& {Butler}}]{feng2017}
{Feng}, F., {Tuomi}, M., {Jones}, H.~R.~A., {et~al.} 2017, \aj, 154, 135

\bibitem[{Feng {et~al.}(2016)Feng, Tuomi, Jones, Butler, \& Vogt}]{FengBIC}
Feng, F., Tuomi, M., Jones, H. R.~A., Butler, R.~P., \& Vogt, S. 2016, Monthly Notices of the Royal Astronomical Society, 461, 2440

\bibitem[{{Figueira} {et~al.}(2010){Figueira}, {Pepe}, {Lovis}, \& {Mayor}}]{HARPS_stability}
{Figueira}, P., {Pepe}, F., {Lovis}, C., \& {Mayor}, M. 2010, \aap, 515, A106

\bibitem[{{Ford} {et~al.}(2001){Ford}, {Havlickova}, \& {Rasio}}]{2001Icar..150..303F}
{Ford}, E.~B., {Havlickova}, M., \& {Rasio}, F.~A. 2001, \icarus, 150, 303

\bibitem[{{Foreman-Mackey} {et~al.}(2013){Foreman-Mackey}, {Hogg}, {Lang}, \& {Goodman}}]{emcee}
{Foreman-Mackey}, D., {Hogg}, D.~W., {Lang}, D., \& {Goodman}, J. 2013, \pasp, 125, 306

\bibitem[{{Ghosh} \& {Chatterjee}(2024)}]{2024MNRAS.527...79G}
{Ghosh}, T. \& {Chatterjee}, S. 2024, \mnras, 527, 79

\bibitem[{{Go{\'z}dziewski} \& {Migaszewski}(2006)}]{HD208487_neptune_14/998}
{Go{\'z}dziewski}, K. \& {Migaszewski}, C. 2006, \aap, 449, 1219

\bibitem[{Gregory(2005)}]{Gregory_longerP}
Gregory, P.~C. 2005, AIP Conference Proceedings, 803, 139

\bibitem[{Gregory(2007)}]{Gregoryupdated}
Gregory, P.~C. 2007, Monthly Notices of the Royal Astronomical Society, 374, 1321

\bibitem[{{Haario} {et~al.}(2006){Haario}, {Laine}, {Mira}, \& {Saksman}}]{haario2006}
{Haario}, H., {Laine}, M., {Mira}, A., \& {Saksman}, E. 2006, Statistics and Computing, 16, 339

\bibitem[{Haario {et~al.}(2001)Haario, Saksman, \& Tamminen}]{haario2001}
Haario, H., Saksman, E., \& Tamminen, J. 2001, Bernoulli, 7, 223

\bibitem[{{Hara} {et~al.}(2020){Hara}, {Bouchy}, {Stalport}, {Boisse}, {Rodrigues}, {Delisle}, {Santerne}, {Henry}, {Arnold}, {Astudillo-Defru}, {Borgniet}, {Bonfils}, {Bourrier}, {Brugger}, {Courcol}, {Dalal}, {Deleuil}, {Delfosse}, {Demangeon}, {D{\'\i}az}, {Dumusque}, {Forveille}, {H{\'e}brard}, {Hobson}, {Kiefer}, {Lopez}, {Mignon}, {Mousis}, {Moutou}, {Pepe}, {Rey}, {Santos}, {S{\'e}gransan}, {Udry}, \& {Wilson}}]{hara2020}
{Hara}, N.~C., {Bouchy}, F., {Stalport}, M., {et~al.} 2020, \aap, 636, L6

\bibitem[{{Hauschildt} {et~al.}(1999){Hauschildt}, {Allard}, \& {Baron}}]{NextGen}
{Hauschildt}, P.~H., {Allard}, F., \& {Baron}, E. 1999, \apj, 512, 377

\bibitem[{{Heller} {et~al.}(2014){Heller}, {Williams}, {Kipping}, {Limbach}, {Turner}, {Greenberg}, {Sasaki}, {Bolmont}, {Grasset}, {Lewis}, {Barnes}, \& {Zuluaga}}]{2014AsBio..14..798H}
{Heller}, R., {Williams}, D., {Kipping}, D., {et~al.} 2014, Astrobiology, 14, 798

\bibitem[{{Henry} {et~al.}(1996){Henry}, {Soderblom}, {Donahue}, \& {Baliunas}}]{HD208487_logrhk}
{Henry}, T.~J., {Soderblom}, D.~R., {Donahue}, R.~A., \& {Baliunas}, S.~L. 1996, \aj, 111, 439

\bibitem[{{Husser} {et~al.}(2013){Husser}, {Wende-von Berg}, {Dreizler}, {Homeier}, {Reiners}, {Barman}, \& {Hauschildt}}]{Pheonix}
{Husser}, T.~O., {Wende-von Berg}, S., {Dreizler}, S., {et~al.} 2013, \aap, 553, A6

\bibitem[{{Izidoro} {et~al.}(2017){Izidoro}, {Ogihara}, {Raymond}, {Morbidelli}, {Pierens}, {Bitsch}, {Cossou}, \& {Hersant}}]{2017MNRAS.470.1750I}
{Izidoro}, A., {Ogihara}, M., {Raymond}, S.~N., {et~al.} 2017, \mnras, 470, 1750

\bibitem[{Jenkins {et~al.}(2020)Jenkins, D{\'\i}az, Kurtovic, Espinoza, Vines, Rojas, Brahm, Torres, Cort{\'e}s-Zuleta, Soto, Lopez, King, Wheatley, Winn, Ciardi, Ricker, Vanderspek, Latham, Seager, Jenkins, Beichman, Bieryla, Burke, Christiansen, Henze, Klaus, McCauliff, Mori, Narita, Nishiumi, Tamura, de~Leon, Quinn, Villase{\~n}or, Vezie, Lissauer, Collins, Collins, Isopi, Mallia, Ercolino, Petrovich, Jord{\'a}n, Acton, Armstrong, Bayliss, Bouchy, Belardi, Bryant, Burleigh, Cabrera, Casewell, Chaushev, Cooke, Eigm{\"u}ller, Erikson, Foxell, G{\"a}nsicke, Gill, Gillen, G{\"u}nther, Goad, Hooton, Jackman, Louden, McCormac, Moyano, Nielsen, Pollacco, Queloz, Rauer, Raynard, Smith, Tilbrook, Titz-Weider, Turner, Udry, Walker, Watson, West, Palle, Ziegler, Law, \& Mann}]{LTT_emp5}
Jenkins, J.~S., D{\'\i}az, M.~R., Kurtovic, N.~T., {et~al.} 2020, Nature Astronomy, 4, 1148

\bibitem[{{Jenkins} {et~al.}(2019{\natexlab{a}}){Jenkins}, {Harrington}, {Challener}, {Kurtovic}, {Ramirez}, {Pe{\~n}a}, {McIntyre}, {Himes}, {Rodr{\'\i}guez}, {Anglada-Escud{\'e}}, {Dreizler}, {Ofir}, {Pe{\~n}a Rojas}, {Ribas}, {Rojo}, {Kipping}, {Butler}, {Amado}, {Rodr{\'\i}guez-L{\'o}pez}, {Kempton}, {Palle}, \& {Murgas}}]{proxcentb_emp2}
{Jenkins}, J.~S., {Harrington}, J., {Challener}, R.~C., {et~al.} 2019{\natexlab{a}}, \mnras, 487, 268

\bibitem[{Jenkins {et~al.}(2025)Jenkins, Jones, Vines, Rubenstein, Peña~Rojas, Wittenmyer, Brahm, Tala~Pinto, \& Carson}]{jenkins2025}
Jenkins, J.~S., Jones, M.~I., Vines, J.~I., {et~al.} 2025, a\&A, submitted

\bibitem[{{Jenkins} {et~al.}(2019{\natexlab{b}}){Jenkins}, {Pozuelos}, {Tuomi}, {Berdi{\~n}as}, {D{\'\i}az}, {Vines}, {Su{\'a}rez}, \& {Pe{\~n}a Rojas}}]{GJ357_emp3}
{Jenkins}, J.~S., {Pozuelos}, F.~J., {Tuomi}, M., {et~al.} 2019{\natexlab{b}}, \mnras, 490, 5585

\bibitem[{{Jenkins} \& {Tuomi}(2014)}]{jenkins2014}
{Jenkins}, J.~S. \& {Tuomi}, M. 2014, \apj, 794, 110

\bibitem[{{Jenkins} {et~al.}(2013){Jenkins}, {Tuomi}, {Brasser}, {Ivanyuk}, \& {Murgas}}]{jenkins2013}
{Jenkins}, J.~S., {Tuomi}, M., {Brasser}, R., {Ivanyuk}, O., \& {Murgas}, F. 2013, \apj, 771, 41

\bibitem[{{Johansen} {et~al.}(2012){Johansen}, {Davies}, {Church}, \& {Holmelin}}]{2012ApJ...758...39J}
{Johansen}, A., {Davies}, M.~B., {Church}, R.~P., \& {Holmelin}, V. 2012, \apj, 758, 39

\bibitem[{Jones {et~al.}(2006)Jones, Sleep, \& Underwood}]{Habitability208487}
Jones, B.~W., Sleep, P.~N., \& Underwood, D.~R. 2006, The Astrophysical Journal, 649, 1010

\bibitem[{{Jones} \& {Jenkins}(2014)}]{jones14}
{Jones}, M.~I. \& {Jenkins}, J.~S. 2014, \aap, 562, A129

\bibitem[{Kass \& Raftery(1995)}]{BayesFactor}
Kass, R.~E. \& Raftery, A.~E. 1995, Journal of the American Statistical Association, 90, 773

\bibitem[{{Kasting} {et~al.}(1993){Kasting}, {Whitmire}, \& {Reynolds}}]{HZMainSequence}
{Kasting}, J.~F., {Whitmire}, D.~P., \& {Reynolds}, R.~T. 1993, \icarus, 101, 108

\bibitem[{{Kipping}(2009)}]{2009MNRAS.392..181K}
{Kipping}, D.~M. 2009, \mnras, 392, 181

\bibitem[{{Kopparapu} {et~al.}(2013){Kopparapu}, {Ramirez}, {Kasting}, {Eymet}, {Robinson}, {Mahadevan}, {Terrien}, {Domagal-Goldman}, {Meadows}, \& {Deshpande}}]{HZ_new}
{Kopparapu}, R.~K., {Ramirez}, R., {Kasting}, J.~F., {et~al.} 2013, \apj, 765, 131

\bibitem[{{Kurucz}(1993)}]{ATLAS9_1993}
{Kurucz}, R. 1993, ATLAS9 Stellar Atmosphere Programs and 2 km/s grid. Kurucz CD-ROM No. 13. Cambridge, 13

\bibitem[{Lissauer(1999)}]{ThreeUpsilonAndromedae}
Lissauer, J.~J. 1999, Nature, 398, 659

\bibitem[{{Lissauer} {et~al.}(2011){Lissauer}, {Ragozzine}, {Fabrycky}, {Steffen}, {Ford}, {Jenkins}, {Shporer}, {Holman}, {Rowe}, {Quintana}, {Batalha}, {Borucki}, {Bryson}, {Caldwell}, {Carter}, {Ciardi}, {Dunham}, {Fortney}, {Gautier}, {Howell}, {Koch}, {Latham}, {Marcy}, {Morehead}, \& {Sasselov}}]{lissauer2011}
{Lissauer}, J.~J., {Ragozzine}, D., {Fabrycky}, D.~C., {et~al.} 2011, \apjs, 197, 8

\bibitem[{{Lissauer} {et~al.}(2024){Lissauer}, {Rowe}, {Jontof-Hutter}, {Fabrycky}, {Ford}, {Ragozzine}, {Steffen}, \& {Nizam}}]{lissauer2024}
{Lissauer}, J.~J., {Rowe}, J.~F., {Jontof-Hutter}, D., {et~al.} 2024, Planetary Science Journal, 5, 152

\bibitem[{{Lo Curto} {et~al.}(2015){Lo Curto}, {Pepe}, {Avila}, {Boffin}, {Bovay}, {Chazelas}, {Coffinet}, {Fleury}, {Hughes}, {Lovis}, {Maire}, {Manescau}, {Pasquini}, {Rihs}, {Sinclaire}, \& {Udry}}]{HARPS_upgrade}
{Lo Curto}, G., {Pepe}, F., {Avila}, G., {et~al.} 2015, The Messenger, 162, 9

\bibitem[{{Lomb}(1976)}]{Lomb}
{Lomb}, N.~R. 1976, \apss, 39, 447

\bibitem[{{Lovis} {et~al.}(2011){Lovis}, {Dumusque}, {Santos}, {Bouchy}, {Mayor}, {Pepe}, {Queloz}, {S{\'e}gransan}, \& {Udry}}]{MagneticCycles_Lovis}
{Lovis}, C., {Dumusque}, X., {Santos}, N.~C., {et~al.} 2011, arXiv e-prints, arXiv:1107.5325

\bibitem[{{Lovis} {et~al.}(2006){Lovis}, {Pepe}, {Bouchy}, {Lo Curto}, {Mayor}, {Pasquini}, {Queloz}, {Rupprecht}, {Udry}, \& {Zucker}}]{HARPS_temp_press}
{Lovis}, C., {Pepe}, F., {Bouchy}, F., {et~al.} 2006, in Society of Photo-Optical Instrumentation Engineers (SPIE) Conference Series, Vol. 6269, Society of Photo-Optical Instrumentation Engineers (SPIE) Conference Series, ed. I.~S. {McLean} \& M.~{Iye}, 62690P

\bibitem[{Marcy \& Butler(1996)}]{70Virginis}
Marcy, G.~W. \& Butler, R.~P. 1996, The Astrophysical Journal, 464, L147

\bibitem[{{Mayor} {et~al.}(2003){Mayor}, {Pepe}, {Queloz}, {Bouchy}, {Rupprecht}, {Lo Curto}, {Avila}, {Benz}, {Bertaux}, {Bonfils}, {Dall}, {Dekker}, {Delabre}, {Eckert}, {Fleury}, {Gilliotte}, {Gojak}, {Guzman}, {Kohler}, {Lizon}, {Longinotti}, {Lovis}, {Megevand}, {Pasquini}, {Reyes}, {Sivan}, {Sosnowska}, {Soto}, {Udry}, {van Kesteren}, {Weber}, \& {Weilenmann}}]{HARPS_original}
{Mayor}, M., {Pepe}, F., {Queloz}, D., {et~al.} 2003, The Messenger, 114, 20

\bibitem[{Mayor \& Queloz(1995)}]{51Peg}
Mayor, M. \& Queloz, D. 1995, Nature, 378, 355

\bibitem[{{Mayor} {et~al.}(2009){Mayor}, {Udry}, {Lovis}, {Pepe}, {Queloz}, {Benz}, {Bertaux}, {Bouchy}, {Mordasini}, \& {Segransan}}]{HD40307_original}
{Mayor}, M., {Udry}, S., {Lovis}, C., {et~al.} 2009, \aap, 493, 639

\bibitem[{{Morton}(2015)}]{Morton_isochrones}
{Morton}, T.~D. 2015, {isochrones: Stellar model grid package}, Astrophysics Source Code Library, record ascl:1503.010

\bibitem[{{Nakajima} {et~al.}(2022){Nakajima}, {Genda}, {Asphaug}, \& {Ida}}]{2022NatCo..13..568N}
{Nakajima}, M., {Genda}, H., {Asphaug}, E., \& {Ida}, S. 2022, Nature Communications, 13, 568

\bibitem[{{Namouni}(2010)}]{2010ApJ...719L.145N}
{Namouni}, F. 2010, \apjl, 719, L145

\bibitem[{{Orell-Miquel} {et~al.}(2023){Orell-Miquel}, {Nowak}, {Murgas}, {Palle}, {Morello}, {Luque}, {Badenas-Agusti}, {Ribas}, {Lafarga}, {Espinoza}, {Morales}, {Zechmeister}, {Alqasim}, {Cochran}, {Gandolfi}, {Goffo}, {Kab{\'a}th}, {Korth}, {Lam}, {Livingston}, {Muresan}, {Persson}, \& {Van Eylen}}]{HD19193_Revisited_HZ}
{Orell-Miquel}, J., {Nowak}, G., {Murgas}, F., {et~al.} 2023, \aap, 669, A40

\bibitem[{{Oshagh} {et~al.}(2017){Oshagh}, {Santos}, {Figueira}, {Barros}, {Donati}, {Adibekyan}, {Faria}, {Watson}, {Cegla}, {Dumusque}, {H{\'e}brard}, {Demangeon}, {Dreizler}, {Boisse}, {Deleuil}, {Bonfils}, {Pepe}, \& {Udry}}]{StellarActivity_RVJitter}
{Oshagh}, M., {Santos}, N.~C., {Figueira}, P., {et~al.} 2017, \aap, 606, A107

\bibitem[{{Petrovich} {et~al.}(2014){Petrovich}, {Tremaine}, \& {Rafikov}}]{2014ApJ...786..101P}
{Petrovich}, C., {Tremaine}, S., \& {Rafikov}, R. 2014, \apj, 786, 101

\bibitem[{Peña~Rojas \& Jenkins(2025)}]{pena2025}
Peña~Rojas, P.~A. \& Jenkins, J.~S. 2025, a\&A, submitted

\bibitem[{{Pojmanski}(1997)}]{Pojmanski_ASAS}
{Pojmanski}, G. 1997, \actaa, 47, 467

\bibitem[{Popova \& Shevchenko(2014)}]{OrbitalResonances}
Popova, E.~A. \& Shevchenko, I.~I. 2014, Journal of Physics: Conference Series, 572, 012006

\bibitem[{Quintana {et~al.}(2014)Quintana, Barclay, Raymond, Rowe, Bolmont, Caldwell, Howell, Kane, Huber, Crepp, Lissauer, Ciardi, Coughlin, Everett, Henze, Horch, Isaacson, Ford, Adams, Still, Hunter, Quarles, \& Selsis}]{Kepler186f}
Quintana, E.~V., Barclay, T., Raymond, S.~N., {et~al.} 2014, Science, 344, 277

\bibitem[{{Ramos} {et~al.}(2017){Ramos}, {Charalambous}, {Ben{\'\i}tez-Llambay}, \& {Beaug{\'e}}}]{Migration_21_32_Resonance}
{Ramos}, X.~S., {Charalambous}, C., {Ben{\'\i}tez-Llambay}, P., \& {Beaug{\'e}}, C. 2017, \aap, 602, A101

\bibitem[{{Rein} \& {Liu}(2012)}]{2012A&A...537A.128R}
{Rein}, H. \& {Liu}, S.~F. 2012, \aap, 537, A128

\bibitem[{Robert(2007)}]{robert_2007}
Robert, C. 2007

\bibitem[{Saha(2022)}]{saha2022precise}
Saha, S. 2022, PhD thesis, Indian Institute of Astrophysics

\bibitem[{{Saha}(2024)}]{2024BSRSL..93..123S}
{Saha}, S. 2024, Bulletin de la Societe Royale des Sciences de Liege, 93, 123

\bibitem[{{Saha} \& {Sengupta}(2022)}]{2022ApJ...936....2S}
{Saha}, S. \& {Sengupta}, S. 2022, \apj, 936, 2

\bibitem[{{Santos} {et~al.}(2001){Santos}, {Mayor}, {Naef}, {Pepe}, {Queloz}, {Udry}, \& {Burnet}}]{HD28185_Discovery}
{Santos}, N.~C., {Mayor}, M., {Naef}, D., {et~al.} 2001, \aap, 379, 999

\bibitem[{{Santos} {et~al.}(2014){Santos}, {Mortier}, {Faria}, {Dumusque}, {Adibekyan}, {Delgado-Mena}, {Figueira}, {Benamati}, {Boisse}, {Cunha}, {Gomes da Silva}, {Lo Curto}, {Lovis}, {Martins}, {Mayor}, {Melo}, {Oshagh}, {Pepe}, {Queloz}, {Santerne}, {S{\'e}gransan}, {Sozzetti}, {Sousa}, \& {Udry}}]{Santos_stellaractivity}
{Santos}, N.~C., {Mortier}, A., {Faria}, J.~P., {et~al.} 2014, \aap, 566, A35

\bibitem[{{Scargle}(1982)}]{Scargle}
{Scargle}, J.~D. 1982, \apj, 263, 835

\bibitem[{{Simon} {et~al.}(2009){Simon}, {Szab{\'o}}, \& {Szatm{\'a}ry}}]{2009EM&P..105..385S}
{Simon}, A.~E., {Szab{\'o}}, G.~M., \& {Szatm{\'a}ry}, K. 2009, Earth Moon and Planets, 105, 385

\bibitem[{{Sneden}(1973)}]{Sneden1973}
{Sneden}, C. 1973, \apj, 184, 839

\bibitem[{{Soto} \& {Jenkins}(2018)}]{SPECIES}
{Soto}, M.~G. \& {Jenkins}, J.~S. 2018, \aap, 615, A76

\bibitem[{{Soto} {et~al.}(2021){Soto}, {Jones}, \& {Jenkins}}]{speciesII}
{Soto}, M.~G., {Jones}, M.~I., \& {Jenkins}, J.~S. 2021, \aap, 647, A157

\bibitem[{{Spalding} {et~al.}(2016){Spalding}, {Batygin}, \& {Adams}}]{2016ApJ...817...18S}
{Spalding}, C., {Batygin}, K., \& {Adams}, F.~C. 2016, \apj, 817, 18

\bibitem[{{Timmermann} {et~al.}(2020){Timmermann}, {Heller}, {Reiners}, \& {Zechmeister}}]{2020A&A...635A..59T}
{Timmermann}, A., {Heller}, R., {Reiners}, A., \& {Zechmeister}, M. 2020, \aap, 635, A59

\bibitem[{{Tinney} {et~al.}(2005){Tinney}, {Butler}, {Marcy}, {Jones}, {Penny}, {McCarthy}, {Carter}, \& {Fischer}}]{TinneyDiscoveryHD208487}
{Tinney}, C.~G., {Butler}, R.~P., {Marcy}, G.~W., {et~al.} 2005, \apj, 623, 1171

\bibitem[{{Trifonov}(2019)}]{exostriker}
{Trifonov}, T. 2019, {The Exo-Striker: Transit and radial velocity interactive fitting tool for orbital analysis and N-body simulations}, Astrophysics Source Code Library, record ascl:1906.004

\bibitem[{{Tuomi} {et~al.}(2013){Tuomi}, {Anglada-Escud{\'e}}, {Gerlach}, {Jones}, {Reiners}, {Rivera}, {Vogt}, \& {Butler}}]{HD40307_reexamine}
{Tuomi}, M., {Anglada-Escud{\'e}}, G., {Gerlach}, E., {et~al.} 2013, \aap, 549, A48

\bibitem[{{Tuomi} {et~al.}(2018){Tuomi}, {Jones}, {Barnes}, {Anglada-Escud{\'e}}, {Butler}, {Kiraga}, \& {Vogt}}]{tuomi18}
{Tuomi}, M., {Jones}, H. R.~A., {Barnes}, J.~R., {et~al.} 2018, \aj, 155, 192

\bibitem[{Tuomi {et~al.}(2014)Tuomi, Jones, Barnes, Anglada-Escud{\'e}, \& Jenkins}]{Tuomi_Bayesian}
Tuomi, M., Jones, H. R.~A., Barnes, J.~R., Anglada-Escud{\'e}, G., \& Jenkins, J.~S. 2014, Monthly Notices of the Royal Astronomical Society, 441, 1545

\bibitem[{{Vines} \& {Jenkins}(2022)}]{ARIADNE}
{Vines}, J.~I. \& {Jenkins}, J.~S. 2022, \mnras, 513, 2719

\bibitem[{Vogt {et~al.}(2010)Vogt, Butler, Rivera, Haghighipour, Henry, \& Williamson}]{Gliese581}
Vogt, S.~S., Butler, R.~P., Rivera, E.~J., {et~al.} 2010, The Astrophysical Journal, 723, 954

\bibitem[{{Vousden} {et~al.}(2016){Vousden}, {Farr}, \& {Mandel}}]{Vousden}
{Vousden}, W.~D., {Farr}, W.~M., \& {Mandel}, I. 2016, \mnras, 455, 1919

\bibitem[{{Weiss} {et~al.}(2018){Weiss}, {Marcy}, {Petigura}, {Fulton}, {Howard}, {Winn}, {Isaacson}, {Morton}, {Hirsch}, {Sinukoff}, {Cumming}, {Hebb}, \& {Cargile}}]{2018AJ....155...48W}
{Weiss}, L.~M., {Marcy}, G.~W., {Petigura}, E.~A., {et~al.} 2018, \aj, 155, 48

\bibitem[{Wittenmyer {et~al.}(2019{\natexlab{a}})Wittenmyer, Bergmann, Horner, Clark, \& Kane}]{Trulyeccentric2}
Wittenmyer, R.~A., Bergmann, C., Horner, J., Clark, J., \& Kane, S.~R. 2019{\natexlab{a}}, Monthly Notices of the Royal Astronomical Society, 484, 4230

\bibitem[{Wittenmyer {et~al.}(2019{\natexlab{b}})Wittenmyer, Clark, Zhao, Horner, Wang, \& Johns}]{Trulyeccentric1}
Wittenmyer, R.~A., Clark, J.~T., Zhao, J., {et~al.} 2019{\natexlab{b}}, Monthly Notices of the Royal Astronomical Society, 484, 5859

\bibitem[{{Wittenmyer} {et~al.}(2017){Wittenmyer}, {Jones}, {Horner}, {Kane}, {Marshall}, {Mustill}, {Jenkins}, {Rojas}, {Zhao}, {Villaver}, {Butler}, \& {Clark}}]{Wittenmyer_astroemperor}
{Wittenmyer}, R.~A., {Jones}, M.~I., {Horner}, J., {et~al.} 2017, \aj, 154, 274

\bibitem[{Wittenmyer {et~al.}(2013)Wittenmyer, Wang, Horner, Tinney, Butler, Jones, O'Toole, Bailey, Carter, Salter, Wright, \& Zhou}]{Testingeccentric_multiples}
Wittenmyer, R.~A., Wang, S., Horner, J., {et~al.} 2013, The Astrophysical Journal Supplement Series, 208, 2

\bibitem[{{Wright} {et~al.}(2007){Wright}, {Marcy}, {Fischer}, {Butler}, {Vogt}, {Tinney}, {Jones}, {Carter}, {Johnson}, {McCarthy}, \& {Apps}}]{HD208487_28d}
{Wright}, J.~T., {Marcy}, G.~W., {Fischer}, D.~A., {et~al.} 2007, \apj, 657, 533

\bibitem[{{Wu} \& {He}(2023)}]{2023AJ....166..267W}
{Wu}, D.-H. \& {He}, Y. 2023, \aj, 166, 267

\bibitem[{{Zechmeister} \& {K{\"u}rster}(2009)}]{GeneralizedLS}
{Zechmeister}, M. \& {K{\"u}rster}, M. 2009, \aap, 496, 577

\bibitem[{{Zhu} {et~al.}(2018){Zhu}, {Petrovich}, {Wu}, {Dong}, \& {Xie}}]{2018ApJ...860..101Z}
{Zhu}, W., {Petrovich}, C., {Wu}, Y., {Dong}, S., \& {Xie}, J. 2018, \apj, 860, 101

\end{thebibliography}


\begin{appendix}

\onecolumn
    \section{RV and stellar activity index data}
    The following tables list the first five RVs and stellar activity index data for each instrument utilized in this paper. The full data can be found at the CDS.

\begin{table}[h]
\caption{\label{tab:full AAT data}First five AAT data}
\centering
\begin{tabular}{lll}
\hline \hline
JD & RV & RV\_error \\
\hline
2451034.17836 & -16.150 & 2.740     \\
2451413.09591 & -22.900 & 2.940     \\
2451683.28594 & -10.290 & 2.810     \\
2451743.20344 & 30.660  & 3.250     \\
2451767.12176 & 18.660  & 2.330     \\
\hline
\end{tabular}
\end{table}

\begin{longtable}{llllll}
\caption{\label{tab:Full TERRA data}First five TERRA data} \\
\hline \hline
JD          & RV      & RV\_error & S-index & FWHM  & BIS    \\
\endfirsthead
\hline
\endhead
2453265.69209 & -22.310 & 0.817     & 0.164   & 8.465 & 22.481 \\
2453269.66244 & -17.369 & 1.281     & 0.163   & 8.468 & 19.289 \\
2453861.92676 & -34.401 & 0.687     & 0.162   & 8.461 & 20.829 \\
2453862.92224 & -34.217 & 0.643     & 0.162   & 8.459 & 17.286 \\
2453863.94041 & -34.169 & 0.634     & 0.163   & 8.466 & 19.206 \\
\hline
\end{longtable}

\begin{longtable}{llll}
\caption{\label{tab:Full PFS data}First five PFS data} \\
\hline \hline
JD          & RV      & RV\_error & S-index \\
\endfirsthead
\endhead
\hline
2455847.60360 & -7.090  & 2.048     & 0.159   \\
2456094.81496 & -15.712 & 1.943     & 0.147   \\
2456141.70123 & 7.073   & 1.676     & 0.142   \\
2456552.63221 & 18.083  & 2.075     & 0.163   \\
2456867.78785 & -20.282 & 1.960     & 0.147   \\ 
\hline
\end{longtable}

\begin{longtable}{llllll}
\caption{\label{tab:Full TERRA2 data}First five TERRA2 data}\\
\hline \hline
JD          & RV      & RV\_error & S-index & FWHM  & BIS    \\
\hline
\endfirsthead
\endhead
2457559.81282    & 9.378   & 1.172     & 0.157   & 8.496 & 30.035 \\
2457559.81885 & 8.200   & 1.215     & 0.157   & 8.497 & 27.733 \\
2457559.82494 & 8.462   & 1.326     & 0.158   & 8.497 & 23.558 \\
2457560.70802 & 8.305   & 0.900     & 0.156   & 8.494 & 25.718 \\
2457560.71411 & 7.898   & 1.010     & 0.157   & 8.498 & 25.647 \\
\hline
\end{longtable}

\newpage

\section{Split RV GLS periodograms}

\begin{figure*}[!htbp]
\centering
\begin{minipage}{0.48\textwidth}
    \includegraphics[width=\linewidth]{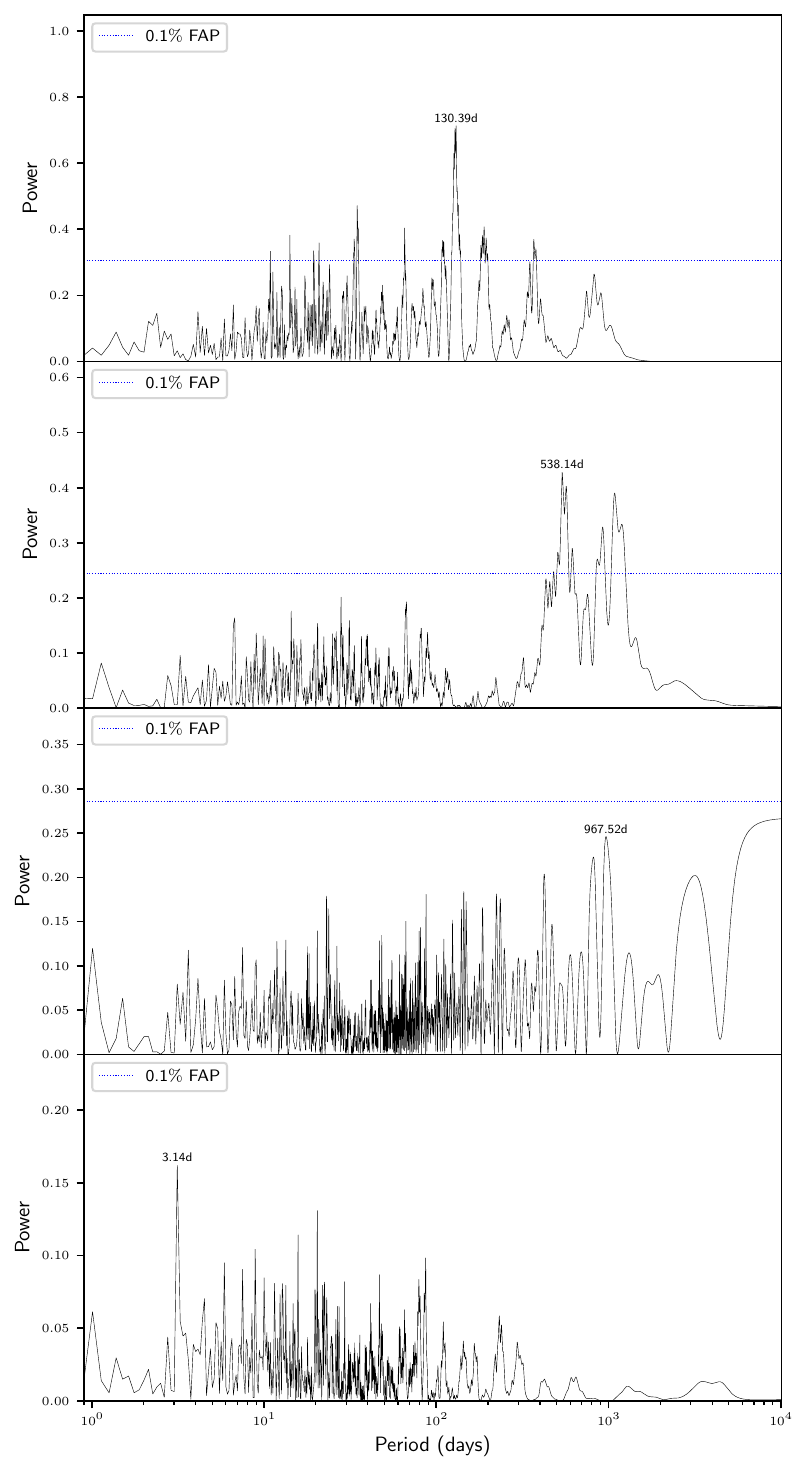}
    \caption{Top panel: original GLS periodogram of the combined and mean-subtracted 
    low S-index RV data. Second panel: GLS periodogram after fitting and removing 
    the first signal. Third panel: GLS periodogram after fitting and removing both 
    the first and second signals. Bottom panel: GLS periodogram after manually 
    fitting the third signal and removing all three previously detected signals. 
    The blue, dotted horizontal lines represent the 0.1\% FAP level estimated from 
    5000 bootstrap resamplings.}
    \label{LowS_Periodograms}
\end{minipage}
\hfill
\begin{minipage}{0.48\textwidth}
    \includegraphics[width=\linewidth]{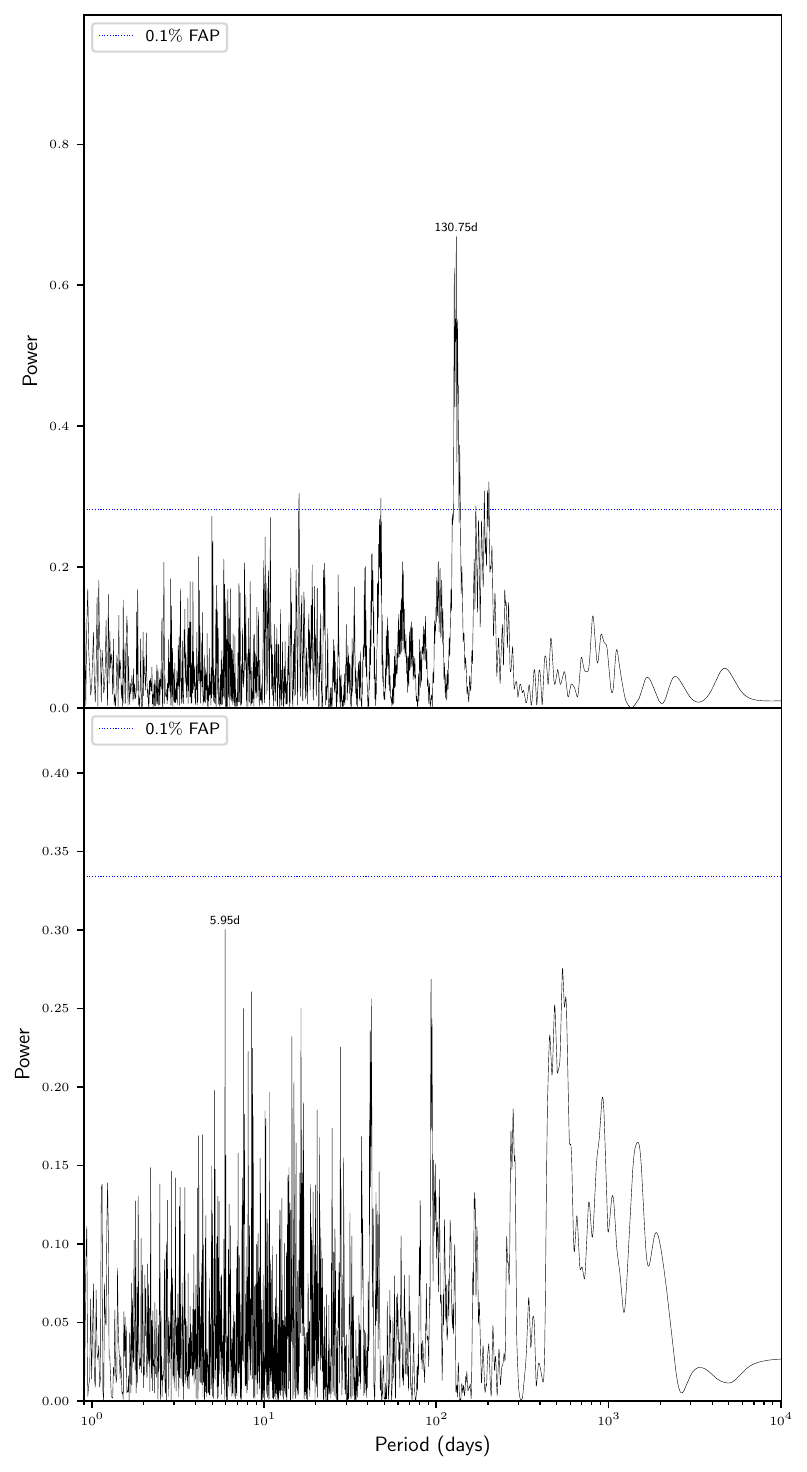}
    \caption{Top panel: original GLS periodogram of the combined and mean-subtracted 
    high S-index RV data. Second panel: GLS periodogram after fitting and removing 
    the first signal. The blue, dotted horizontal lines represent the 0.1\% FAP level 
    estimated from 5000 bootstrap resamplings.}
    \label{HighS_Periodograms}
\end{minipage}
\end{figure*}

Figure~\ref{LowS_Periodograms} shows the GLS Periodogram of the Low S-index dataset, 
while Figure~\ref{HighS_Periodograms} shows the GLS Periodogram of the High S-index dataset.

\end{appendix}

\end{document}